\documentclass[footinbib,prx,preprint]{revtex4-2}
\usepackage{graphicx}% Include figure files
\usepackage{dcolumn}% Align table columns on decimal point
\usepackage{bm}% bold math
\usepackage{color}
\usepackage{soul}

\usepackage{amsmath,amsfonts,amssymb}
\usepackage{physics}

\newcommand{\nc}{\newcommand}
\nc{\nn}{\nonumber}
\def\beq{\begin{equation}}
\def\eeq{\end{equation}}
\def\beqa{\begin{eqnarray}}
\def\eeqa{\end{eqnarray}}
\def\e{\mathcal{E}}

\nc{\XYZ }{}

\begin{document}
\title{% A comprehensive electronic and thermal theory of the response of ITO to ultrafast illumination
The electronic and thermal response of low electron density Drude materials to ultrafast optical illumination}

\date{\today}

\author{Subhajit Sarkar}
\email{subhajit@post.bgu.ac.il}
\thanks{These two authors have contributed equally.}
\author{Ieng Wai Un}
\email{iengwai@post.bgu.ac.il}
\thanks{These two authors have contributed equally.}
\author{Yonatan Sivan}
\email{sivanyon@bgu.ac.il}
\affiliation{School of Electrical and Computer Engineering, Ben-Gurion University of the Negev and the Ilse Katz Center for Nanoscale Science and Technology, Ben-Gurion University of the Negev, Beer Sheva, Israel}

% $^\dagger$ - equal contribution

\begin{abstract}
Many low electron density Drude (LEDD) materials such as transparent conductive oxide or nitrides have recently attracted interest as alternative plasmonic materials and future nonlinear optical materials. However, the rapidly growing number of experimental studies has so far not been supported by a systematic theory of the electronic, thermal and optical response of these materials. Here, we use the techniques previously derived in the context of noble metals to go beyond a simple electromagnetic modelling of low electron density Drude materials and provide an electron dynamics model for their electronic and thermal response. % In particular, we compute the electron dynamics without resorting to the RTA within the Thomas-Fermi approach, and treat the phonon system self-consistently.
We find that the low electron density makes momentum conservation in $e-ph$ interactions more important, more complex and more sensitive to the temperatures compared with noble metals; moreover, we find that $e-e$ interactions are becoming more effective due to the weaker screening. Most importantly, we show that the low electron density makes the electron heat capacity much smaller than in noble metals, such that the electrons in LEDD materials tend to heat up much more and cool down faster compared to noble metals.
%As a consequence of the intense illumination and associated high electron temperature, the effective chemical potential dramatically decreases and even becomes negative, thus, effectively converting the Drude metal into a semiconductor.
While here we focus on indium tin oxide (ITO), our analytic results can be easily applied to any LEDD materials. 
\end{abstract}

\maketitle

\section{Introduction}
Recent years have seen a growing interest in the optical response of plasmonic (transparent) conducting oxides and nitrides such as aluminum/gallium-doped zinc oxide, cadmium oxide, indium tin oxide, titanium/zirconium nitride etc.~\cite{Guru_Boltasseva_alternatives,Boyd_NLO_ENZ_ITO,Shalaev_Schaller_adom,Yang_polarization_switching_NP_2017,Khurgin_ENZs_nlty,Shalaev_Faccio_NLO_ENZ} as alternatives to noble metals. These materials are characterized by low electron densities and high frequency interband transitions, such that they are usually described as Drude metals at optical frequencies~\cite{Guru_Boltasseva_alternatives}. In comparison to noble metals, the real part of their permittivities has a milder negative value, the electron density (hence, permittivity) is highly tunable (e.g., via doping~\cite{Guru_Boltasseva_alternatives,Feigenbaum-Atwater-doping} or post-treatment etc.~\cite{Liu_ITO_2014,pradhan2014extreme,chinese_superfluid_treatment_ITO}), so that together with their CMOS compatibility, they enable flexibility of design and implementation~\cite{Guru_Boltasseva_alternatives}. In what follows, we refer to this class of materials as low electron density Drude (LEDD) materials.

Particular attention has been given to the unique epsilon-near-zero (ENZ) point these materials have in the near infrared spectral range~\cite{Boyd_NLO_ENZ_ITO,Boyd_Nat_Phot_2018,Khurgin_ENZs_nlty,Exeter_Nat_Comm_2021}. % Particular attention is given to the popular transparent conducting oxide ITO which emerges as an interesting candidate for high efficiency ultrafast nonlinear optics applications. 
While most earlier interest in ENZ materials was associated with their linear response (e.g., in the context of supercoupling~\cite{Engheta-supercoupling,Smith-ENZ} or shaping the radiation pattern of a source~\cite{ENZ_Engheta,ENZ_Engheta_Polman}), the realization that the nonlinear optical response is inversely proportional to the (unperturbed vanishing) permittivity attracted a range of experimental studies of the (ultrafast) dynamics of the permittivity near the ENZ point (e.g.,~\cite{Boyd_NLO_ENZ_ITO,Guo_ITO_nanorods_NC_2016,Guo_ITO_AM_2017,Boyd_Nat_Phot_2018,Yang_polarization_switching_NP_2017,Exeter_Nat_Comm_2021,Sapienza_2022}). In parallel, the importance of the deviation of the band structure from a simple parabolic dependence was realized~\cite{Liu_ITO_2014,Yang_polarization_switching_NP_2017,Shalaev_Schaller_adom}. From the theoretical point of view, most attention was given to modelling the dependence of the LEDD permittivity on the electron temperature using thermal models~\cite{Guo_ITO_nanorod_natphoton,Guo_ITO_nanorods_NC_2016,Yang_polarization_switching_NP_2017,Guo_ITO_nanorod_natphoton,Xian_group_ITO_2019} while the temperature dynamics itself was described using the Relaxation Time Approximation (RTA)-based Two Temperature Model (TTM) or its extension~\cite{Guo_ITO_nanorod_natphoton,Guo_ITO_nanorods_NC_2016,Yang_polarization_switching_NP_2017,Boyd_Nat_Phot_2018,Guo_ITO_nanorod_natphoton,Guo_ITO_AM_2017,Shalaev_Schaller_adom}. However, sometimes the TTM parameters were computed from equations suitable for parabolic bands and high density metals. In that regard, the accuracy of the temperature modelling in LEDD materials has not yet been verified using an underlying electronic model.

Our goal in the current work is to compute consistently the non-equilibrium electron dynamics in LEDD materials. We employ the techniques previously used in the context of noble metals to go beyond the state-of-the-art modelling of LEDD materials and provide a model for their electronic and thermal dynamical response. We focus on indium tin oxide (ITO) as a prototypical example. In particular, in Section~\ref{sec:BE}, we derive the various terms in the simplest model available for electron non-equilibrium, namely, the Boltzmann equation (BE) without resorting to the Relaxation Time Approximation (RTA); we complement the BE with a self-consistent treatment of the phonon system. We find that due to the low electron density, hence, weaker screening, the $e-e$ collisions are 10 times faster than in the higher electron density noble metals. We also find that for the same reason, the requirement of momentum conservation in $e-ph$ interactions significantly slows down the collision rate; however, due to the opposing effect of the higher Debye energy, the $e-ph$ collision rate is actually similar to that in noble metals. In Section~\ref{subsec:e-dynamics} we describe the resulting ultrafast dynamics of the electron distribution, and correlate it with the relative importance of the various underlying collision mechanisms. In Section~\ref{subsec:TTM} we derive analytic expressions for the TTM parameters and find that the dependence of both heat capacity and chemical potential on the electron temperature is much stronger than assumed so far, and their values are much lower compared to noble metals. We then extract the temperature dynamics from the BE and show an excellent match with the TTM. Our main results are that the electron heating is much stronger and its decay is much faster in ITO compared to noble metals due to the lower electron heat capacity.
%Moreover, we show that due to the low electron density and small Fermi energy, a sufficiently high intensity illumination can heat up the electrons even up to the Fermi temperature, in which case the chemical potential become negative. Under these conditions, the LEDD material effectively becomes a semiconductor.
In Section~\ref{sec:discussion} we conclude with further comparison to more advanced theoretical approaches, and with a discussion of future steps.

\section{Model for the electron distribution}\label{sec:BE}
\begin{figure}[h]
\centering
\includegraphics[width=1\textwidth]{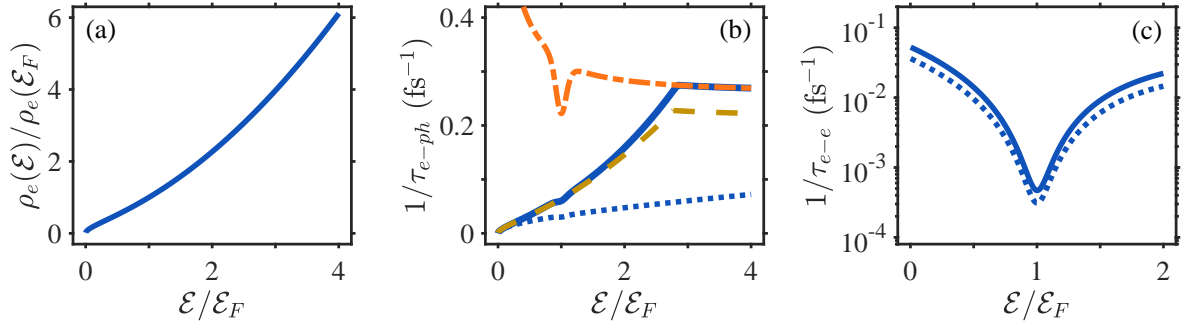}
\caption{(Color online) (a) The normalized electron density of states ${\rho_e(\e)}/{\rho_e(\e_F)}$ as a function of $\e/\e_F$. (b) The collision rate associated with the $e-ph$ interaction ${1}/{\tau_{e-ph}}$~\eqref{eq:invtau_ephn} with (blue solid line) and without (orange dash-dotted line) accounting for momentum conservation and non-parabolicity (i.e., $C\neq 0$). The blue dotted line represents the case accounting for momentum conservation but not for non-parabolicity (i.e., $C = 0$). The dark yellow dashed line represents the approximation of the $e-ph$ collision rate~\eqref{eq:invtau_ephn_simp}. (c) The thermalization rate associated with the $e-e$ interaction with (blue solid line, $C \neq 0$) and without non-parabolicity (blue dotted line, $C = 0$). The collision rates in (b) and (c) are calculated at $T_e = 500$ K and $T_{ph} = 300$ K.}
\label{fig:dos_invtau}
\end{figure}

We determine the electron distribution in LEDD materials by solving the quantum-like Boltzmann equation (BE). This model is in wide use for Drude metals~\cite{Ziman-book,Ashcroft-Mermin,Lundstrom-book,non_eq_model_Lagendijk,delFatti_nonequilib_2000,vallee_nonequilib_2003,Italians_hot_es,Seidman-Nitzan-non-thermal-population-model,GdA_hot_es}. We focus on Indium Tin Oxide (ITO) because it has all the unique features of a LEDD material, namely, the low electron density, a non-parabolic conduction band and because it emerges as a promising CMOS compatible nonlinear optical material. 
% It is valid for nanoparticles which are more than a few nm in size (hence, not requiring energy discretization)~\cite{GdA_hot_es,Govorov_ACS_phot_2017} and for systems where coherence and correlations between electrons are negligible. The latter assumption holds for a simple metal at room temperatures (or higher), as it has a large density of electrons and fast collision mechanisms. In order to include quantum finite size effects or quantum coherence effects, one can use the known relation between the discretized BE and quantum master equations~\cite{Goodnick_DM_BE,Chattah_DM_BE} or by replacing the BE by that equation~\cite{Govorov_1,Govorov_2,Govorov_ACS_phot_2017}.
The energy-momentum ($\e-k$) relation of ITO can be expressed by the Kane quasi-linear dispersion~\cite{Kane-quasilinear} to account for the nonparabolicity~\cite{Guo_ITO_nanorod_natphoton,ITO_Nonparabolicity}, namely,
\begin{align}\label{eq:ITO-E-k}
\dfrac{\hbar^2 k^2}{2 m_e^{\ast}} = \e(1 + C\e),
\end{align}
where $m_e^{\ast} = 0.3964 m_e$ is the electron effective mass at the conduction band minimum, and $C = 0.4191$eV\textsuperscript{-1}~\cite{Liu_ITO_2014} is the first-order nonparabolicity factor. The electron density of states (eDOS) becomes
\begin{align}\label{eq:DOS}
\rho_e(\e) = \dfrac{1 + 2C\e}{2\pi^2}\left(\dfrac{2m_e^\ast}{\hbar^2}\right)^{3/2} \sqrt{\e(1 + C\e)}.
\end{align}
Compared with the case in which $C = 0$ (no non-parabolicity), the density of states when $C \neq 0$ is a superlinear function (instead of a radical function) of the electron energy and is larger by a factor of $\sqrt{1 + C \e}(1 + 2 C \e)$, see Fig.~\ref{fig:dos_invtau}(a). The $\e-k$ relation~(\ref{eq:ITO-E-k}) allows us to represent the electron states in terms of energy $\e$ rather than momentum. We also neglect interband ($d$ to $sp$) transitions which occur in ITO only for photons having energies larger than $3-3.5$eV~\cite{Franzen_ITO_vs_AuAg}; these transitions, however, set the value of the background permittivity, $\varepsilon_b$. 

% , as noted in~\cite{Govorov_ACS_phot_2017}, interband transitions are not likely to generate electrons with energies far above the Fermi level unless the photon energy is much higher than the bandgap energy.

The Boltzmann equation representing electron dynamics is
% \begin{eqnarray}\label{eq:f_neq_dynamics}
% \frac{\partial f\left(\e(\vec{k});T_e,T_{ph}\right)}{\partial t} &=& \underbrace{\left(\frac{\partial f}{\partial t}\right)_{ex}}_{photon\ absorption} + \underbrace{\left(\frac{\partial f}{\partial t}\right)_{e-ph}}_{e-ph\ collisions} + \underbrace{\left(\frac{\partial f}{\partial t}\right)_{e-e}}_{e-e\ collisions},
% \end{eqnarray}
\begin{eqnarray}\label{eq:f_neq_dynamics}
\frac{\partial f}{\partial t} &=& \left(\frac{\partial f}{\partial t}\right)_{exc} + \left(\frac{\partial f}{\partial t}\right)_{e-ph\ collision} + \left(\frac{\partial f}{\partial t}\right)_{e-e\ collision}, % + \left(\frac{\partial f}{\partial t}\right)_{e-imp\ collision},
\end{eqnarray}
where $f(\e)$ is the electron distribution function at an energy $\e$, representing the population probability of electrons in a system characterized by a continuum of states within the conduction band; this description was shown to be suitable for particles of Drude metals of sizes as small as a few nm~\cite{Khurgin-Levy-ACS-Photonics-2020}. 

The first term on the right-hand-side (RHS) of Eq.~(\ref{eq:f_neq_dynamics}) describes excitation of conduction electrons due to photon absorption, see Section~\ref{sub:QM_absorption} below for its explicit form. The second term on the RHS of Eq.~(\ref{eq:f_neq_dynamics}) describes the population relaxation due to collisions between electrons and phonons, see Section~\ref{sub:e-ph} below for its explicit form. %This interaction makes the electrons in our model only quasi-free. 
The third term on the RHS of Eq.~(\ref{eq:f_neq_dynamics}) (see Section~\ref{sub:e-e} below for its explicit form) represents implicitly the thermalization induced by electron-electron ($e-e$) collisions, i.e., the convergence of the {\em non}-thermal population into the {\em thermalized} Fermi-Dirac distribution, given by
\begin{equation}\label{eq:f_eq}
f^T\left(\e;T_e\right) = \left(1 + e^{(\e - \mu)/k_B T_e}\right)^{-1},
\end{equation}
where $\mu$ is the chemical potential, $k_B$ is the Boltzmann constant, and $T_e$ is the electron temperature. Note that collisions of conduction electrons with impurities may also be included, but while those contribute to the permittivity, they have no effect on the electron distribution itself (on the level of the BE~(\ref{eq:f_neq_dynamics}))~\cite{Ashcroft-Mermin}. % The last term on the right-hand-side (RHS) of Eq.~(\ref{eq:f_neq_dynamics}) describes collisions of conduction electrons with impurities, see SI Section~\ref{sub:e-imp} below for its explicit form. 

% let us not relate to this - Within this description, it was shown in~\cite{hot_es_Atwater,GdA_hot_es} that the total electron collision time is independent of the dimensions of the nanostructure~\footnote{Notably, this is in contrast to the claims in~\cite{hot_es_review_2015} which were not supported by evidence. }. 

Our model does not account for electron acceleration due to the force exerted on them by the electric field (which involves a classical description, see discussion in~\cite{non_eq_model_Rethfeld,Dubi-Sivan}) nor for drift due to its gradients or due to temperature gradients; these effects may be important only for nanostructures with complex geometries.
% Further, as customary, we neglect the (weaker) Lorentz force and quantum pressure (which usually appears in the hydrodynamic model for the nonlocal and/or second-order nonlinearity of metals~\cite{scalora_model_SHG_THG}), and neglect higher-order processes (involving more than one photon) which are assumed to be much less probable.
Similar simplifications were adopted in most previous studies of electron non-equilibrium in Drude metals, e.g.,~\cite{GdA_hot_es,Govorov_ACS_phot_2017,Dubi-Sivan}. % These neglected effects can be implemented in our formalism in a straightforward way.

% \begin{table}\label{table:params}\centering
% \caption{Parameters used in the simulations; values chosen for ?? }
% \begin{tabular}{|l|c|c|} \hline
% parameter & parameter symbol & value \\
% \hline
%\midrule
% photon wavelength & $\lambda$ & $1$eV; unnecessary? \\ \hline
% permittivity & $\epsilon_{ITO}(\lambda)$ & $-?? + 0.37??i$~\cite{Indian_Ag_ellipsometry_2014} \\ \hline
% Fermi energy & $\e_F$ & $0.88$eV \\ \hline
% conduction band width & $\e_{max}$ & $?$eV \\ \hline - better not mention this, because we are twicking it...
% chemical potential & $\mu$ & $??$eV; $I$ dependent... \\ \hline
% ph-env coupling & $G_{ph-env}$ & $5 \cdot 10^{14} W/m^3 K$ \\ \hline
% electron density & $n_e$ & $1.5 \cdot 10^{27} m^{-3}$ \\ \hline
% speed of sound & $v_{ph}$ & 6400 m/s; mention in text? \\ \hline
% environment temperature & $T_{env}$ & $297 K$; unnecessary? \\ \hline
% mass density & $\rho$ & $0.5 g/cm^3$ - removed on purpose.. \\ \hline
% particle radius WHERE ASSUMED?? & $a$ & $5$ nm \\ \hline
% effective electron mass & $m_e^*$ & $m_e^{\ast} = 0.3964 m_e$ \\ \hline
%\bottomrule
% \end{tabular}
% \label{tab:params}
% \end{table}

Below, we set the electron density in ITO to be $n_e \sim 1.5 \times 10^{27}$ m\textsuperscript{-3}, such that it is characterized by a relatively low Fermi energy of $\e_F = 0.88$ eV (compared to noble metals) and a total size of $\e_\text{max} = 4$ eV~\cite{ITO_band_structure_2001,ITO_properties_JPCM}. However, note that due to the non-stochiometric nature of ITO (i.e., the dependence on its preparation conditions~\cite{Liu_ITO_2014,ITO_properties_EBV}), the electron density in ITO can vary from $10^{27}$ m\textsuperscript{-3} to $10^{28}$ m\textsuperscript{-3}, %one should expect variations in this value, 
see~\cite{ITO_properties_JPCM} and references therein. Similar and even lower values are typical of other LEDD materials~\cite{Guru_Boltasseva_alternatives}.

Similarly, the values of other parameters indicated below such as the deformation potential, the sound velocity, Debye temperature etc. or even the impurity density or grain size may also vary significantly between sample to sample. Yet, the analysis presented below remains generic to all ITO and other LEDD materials.

\subsection{The quantum mechanical excitation term}\label{sub:QM_absorption}
%We consider the incident pulsed illumination having an electric field described as a Gaussian envelope superimposed on a carrier wave ${\bf E}_\text{inc}(t)= {\bf E}_0e^{-2\ln 2 (t/\tau_\text{pump})^2} e^{-i\omega_\text{pump}t}$, where $\tau_\text{pump}$ is the pulse duration and $\omega_\text{pump}$ is the carrier angular frequency. For simplicity, we further assume that the pulse duration is long enough $\omega_\text{pump}\tau_\text{pump} \gg 1$ such that the incident pulse can be considered to be quasi-monochromatic {\bf why is this necessary?}. 

The absorption of the incident light at frequency $\omega$ leads to the excitation of electrons with initial energy $\e_\text{initial} = \e$ to states with final energy $\e_\text{final} = \e + \hbar\omega$. Here, we employ the expression proposed in~\cite{Seidman-Nitzan-non-thermal-population-model,Dubi-Sivan} to link this process to the local electric field. Specifically, the excitation term can be written as~\cite{Seidman-Nitzan-non-thermal-population-model,delFatti_nonequilib_2000,Dubi-Sivan}
\begin{multline}\label{eq:dfEdt_exc}
\left(\dfrac{\partial f(\e)}{\partial t}\right)_\text{exc} = A(t)\Big[D_J(\e - \hbar\omega,\e)\rho_e(\e-\hbar\omega)f(\e - \hbar\omega)(1 - f(\e)) \\
- D_J(\e,\e + \hbar\omega)\rho_e(\e + \hbar\omega) f(\e)(1 - f(\e + \hbar\omega))\Big].
\end{multline}
Here, $D_J(\e_\text{initial},\e_\text{final})$ is the squared magnitude of the transition matrix element for the electronic process $\e_\text{initial} \rightarrow \e_\text{final}$ (be it Landau damping, surface/phonon-assisted absorption, etc.~\cite{hot_es_Atwater,Khurgin_Landau_damping,GdA_hot_es})% Note that our model does not require indicating what is the  but rather, it accounts only for their cumulative rate. 
; $A(t)$ is a time-dependent coefficient ensuring energy conservation that is proportional to the local energy density, or $|{\bf E}(t)|^2$ (${\bf E}(t)$ being the {\em local} electric field vector in the ITO sample)\footnote{For simplicity, we neglect any inhomogeneity of the local electric field in the ITO sample. }% hence, use of $||$ for its norm)
; in particular, the increase of the energy of the electron system is equal to the energy of the absorbed pulse (via Poynting's Theorem)
\begin{equation}\label{eq:P_abs}
P_\text{abs} \equiv \int \e \rho_e(\e) \left(\dfrac{\partial f(\e)}{\partial t}\right)_\text{exc} d\e = \dfrac{\omega}{2} \varepsilon_0 \varepsilon^{\prime\prime}|{\bf E}(t)|^2,
\end{equation}
so that
\begin{equation}\label{eq:A}
A(t) = \dfrac{\varepsilon_0\varepsilon^{\prime\prime}|{\bf E}(t)|^2}{2\hbar\int D_J(\e-\hbar\omega,\e) \rho_e(\e)\rho_e(\e - \hbar\omega)f(\e - \hbar\omega,t)(1 - f(\e,t))d\e},
\end{equation}
where $\varepsilon^{\prime\prime}$ is the imaginary part of the ITO permittivity.

% The only term computed in energy space? justify the validity of this approach - for particles whose size is not smaller than a few nm (see discussion in~\cite{Khurgin-Levy-ACS-Photonics-2020,Dubi-Sivan})? 

\subsection{The $e-ph$ collision term}\label{sub:e-ph}

%{\bf so the following is gonna be removed?} The $e-ph$ collision term is derived from the Hamiltonian describing electron-phonon interaction via the deformation potential collision, namely~\cite{Snoke_solid_state},
%\begin{align}
%\mathcal{H}_\text{e-ph} = \sum_{{\bf k},{\bf q}}D q \sqrt{\dfrac{\hbar}{2\rho V \omega_{ph}^{{\bf q}}}} i\left(a_{\bf q}c^{\dagger}_{{\bf k} + {\bf q}}c_{{\bf k}} - a^{\dagger}_{\bf q}c^{\dagger}_{{\bf k}}c_{{\bf k} + {\bf q}}\right),
%\end{align}
%where $D$ is the deformation potential, $\rho$ is the material density, $\omega_{\bf q}$ is the angular frequency of the phonon and $a^\dagger_{{\bf q}}$ ($a_{{\bf q}}$) is the creation (annihilation) operator of a phonon with momentum ${\bf q}$. 
The collision term due to electron-phonon interaction via the deformation potential scattering is given by~\cite{Snoke_solid_state},
\begin{align}\label{eq:dfkdt_ephn}
\left(\dfrac{df_{{\bf k}}}{dt}\right)_{e-ph} = \sum_{{\bf q}}&\dfrac{\pi D^2 q^2}{\rho V \omega_{ph}^{{\bf q}}} \Bigg\{\Big[(n^{{\bf q}}_{ph} + 1)(1 - f_{{\bf k}}) f_{{\bf k} + {\bf q}} - n^{{\bf q}}_{ph}(1 - f_{{\bf k} + {\bf q}}) f_{{\bf k}}\Big] \delta(\e_{{\bf k} + {\bf q}} - \e_{{\bf k}} - \hbar \omega_{ph}^{{\bf q}}) \nn \\
-&\ \Big[(n^{{\bf q}}_{ph} + 1)(1 - f_{{\bf k} - {\bf q}}) f_{{\bf k}} - n^{{\bf q}}_{ph} (1 - f_{{\bf k}}) f_{{\bf k} - {\bf q}}\Big] \delta(\e_{{\bf k} - {\bf q}} - \e_{{\bf k}} + \hbar \omega_{ph}^{{\bf q}})\Bigg\},
\end{align}
where ${\bf k}$ is the electron momentum, ${\bf q}$ is the phonon momentum, $\hbar\omega_{ph}^{{\bf q}}$ is the phonon energy, $\rho = 7120$ kg/m\textsuperscript{3} is the material density~\cite{ITO_properties_2016}, $D = 17.2$ eV is the deformation potential~\footnote{This value of the deformation potential was obtained by fitting the ITO permittivity (at room temperature) calculated by the Lindhard formula~\cite{Grosso_solid_state} with that measured experimentally~\cite{Xian_group_ITO_2019}.} and $n^{{\bf q}}_{ph}$ is the phonon distribution function.

For simplicity, we consider only acoustic phonons since they have been found to be responsible for the dominant scattering mechanism compared to optical phonons~\cite{Xian_group_ITO_2020}. We also assume that the acoustic phonons propagate at the sound velocity such that they have a linear dispersion relation, i.e., $\e_{ph}^{{\bf q}} = \hbar\omega_{ph}^{{\bf q}} = \hbar v_{ph} q$ where $v_{ph} = 6400$ m/s~\cite{ITO_properties_2016}. Beyond the Debye energy, $\e_D = k_B T_D \approx 0.06$eV~\cite{ITO_properties_2016}, the density of phonon states vanishes. %Following the procedure in Section 4.8 of~\cite{Snoke_solid_state} based on the Fermi’s golden rule, we obtain the change rate of the electron distribution $$\partial f/\partial t$. 
Furthermore, we have assumed that the phonon system is in equilibrium, so that the average phonon number satisfies the Bose–Einstein statistics and can be characterized by the phonon (lattice) temperature $T_{ph}$,  i.e., $n^{{\bf q}}_{ph}(T_{ph}) = n_B(\e_{ph}^{{\bf q}},T_{ph}) = \left(e^{\hbar\omega^{{\bf q}}_{ph}/k_B T_{ph}} - 1\right)^{-1}$. %, where $T_{ph}$ is the phonon temperature (temperature of the lattice).
The two terms associated with $(n_{ph}^{{\bf q}} + 1)$ correspond to the phonon emission, whereas the two terms associated with $n_{ph}^{{\bf q}}$ correspond to the phonon absorption. The delta-functions in Eq.~\eqref{eq:dfkdt_ephn} ensures energy conservation in the various electron-phonon scattering processes. 

To obtain the collision term in terms of the electron energy $\left(\dfrac{\partial f(\e)}{\partial t}\right)_{e-ph}$, we perform the spherical average over the solid angle for $\left(\dfrac{df_{{\bf k}}}{dt}\right)_{e-ph}$ in Eq.~\eqref{eq:dfkdt_ephn}, namely, $\left(\dfrac{\partial f(\e)}{\partial t}\right)_{e-ph} = \dfrac{1}{4\pi}\displaystyle\int \left(\dfrac{\partial f_{{\bf k}}}{\partial t}\right)_{e-ph} d\cos\theta_{{\bf k}} d\phi_{{\bf k}}$, where $\theta_{{\bf k}}$ and $\phi_{{\bf k}}$ are, respectively, the polar and azimuthal angles of ${\bf k}$ with respect to ${\bf q}$. In particular, we account for the phonon dispersion in the polar integral to ensure momentum conservation, 
\begin{align}\label{eq:eph_polar_int}
\int_{-1}^1 d\cos\theta_{{\bf k}}\delta(\e_{{\bf k} \pm {\bf q}} - \e_{{\bf k}}\mp\hbar\omega_{ph}^{{\bf q}}) = \begin{cases}
\dfrac{m_e^\ast}{\hbar^2 kq},& \text{if }\left|\dfrac{m_e^\ast v_{ph}}{\hbar k}\mp\dfrac{q}{2k}\right|\leq 1, \\
0,&\text{otherwise}
\end{cases}.
\end{align}
Since $m_e^\ast v_{ph} \ll \hbar q_D = \hbar \e_D/v_{ph}$, %{\bf IW/SS - what is the meaning of $m_e^\ast v_{ph}$?? it mixes electron and phonon aspects...: I guess this is standard, e.g., similar condition appear in Ridley's book too.}
the condition for the polar integral being non-zero in Eq.~\eqref{eq:eph_polar_int} can be simplified to $q \lesssim 2k$. %(i.e., $\e_{ph} \lesssim \sqrt{8 m_e^\ast v_{ph}^2\e(1 + C\e)}$). 
After some algebra, we arrive at 
\begin{multline}\label{eq:dfEdt_eph}
\left(\dfrac{\partial f(\e)}{\partial t}\right)_{e-ph} = \ \dfrac{D^2}{4\pi\rho (\hbar v_{ph})^4} \sqrt{\dfrac{m_e^\ast}{2\e(1 + C\e)}}\int_0^{\e_{ph}^\text{max}(\e)}(\e_{ph})^2 d\e_{ph} \\ 
\Big[(n_B(\e_{ph},T_{ph}) + 1)\Big((1 - f(\e)) f(\e + \e_{ph})(1 + 2C(\e + \e_{ph})) \\
- (1 - f(\e - \e_{ph}))f(\e)(1+2C(\e - \e_{ph}))\Big) \\
+ n_B(\e_{ph},T_{ph}) \Big((1 - f(\e))f(\e - \e_{ph})(1 + 2 C (\e - \e_{ph})) \\
- (1 - f(\e + \e_{ph})f(\e)(1 + 2 C(\e + \e_{ph}))\Big)\Big],
\end{multline}
where
%\begin{align}\label{eq:eng_phn_max}
%\e_{ph}^\text{max}(\e) &= \min(\e_D,\sqrt{8 m_e^\ast v_{ph}^2 \e(1+C\e)})= \begin{cases}
%\sqrt{8 m_e^\ast v_{ph}^2 \e(1 + C \e)}) &\text{for } \e \leq \e_\text{min,D} \\
%\e_D & \text{for }\e > \e_\text{min,D}
%\end{cases},
%\end{align}
\begin{align}\label{eq:eng_phn_max}
\e_{ph}^\text{max}(\e) = \text{min}(\e_D,2\hbar v_{ph}k(\e)) = 
\begin{cases}
2\hbar v_{ph}k(\e)&\text{for } k(\e)< q_D/2\\
\e_D\equiv\hbar v_{ph} q_D&\text{for } k(\e)\geq q_D/2
\end{cases}
\end{align}
is the maximal energy of a phonon which can be absorbed/emitted by an electron with energy $\e$~\cite{coh_ele_dyn_thm_latt_vib,Ridley_QM_process_SC_2013}, $\hbar q_D = \e_D/v_{ph}$ is the Debye momentum and $k(\e)$ is the momentum of an electron with energy $\e$ obtained from the dispersion relation~\eqref{eq:ITO-E-k}. %This means that only the phonons with energy smaller than $\e_{ph}_\text{max}(\e)$ contribute to the $e-ph$ scattering because of momentum conservation. 
%Here, we defined $\e_\text{min,D} \equiv \dfrac{1}{2C}\left(\sqrt{1 + \dfrac{C\e_D^2}{2 m_e^\ast v_{ph}^2}} - 1\right)$ such that only those electrons having energy higher than $\e_\text{min,D}$ can interact with all phonons, while for electrons having energy less than $\e_\text{min,D}$, the number of phonons available for $e-ph$ collisions is suppressed due to momentum conservation. 
%This can be further understood using a phase-space argument, as detailed in Appendix~\ref{app:e-ph_phase_space}. 

By taking the functional derivative of Eq.~\eqref{eq:dfEdt_eph} with respect to $f(\e)$~\cite{Quantum-Liquid-Coleman}, we obtain the collision rate associated with the $e-ph$ interaction, $\dfrac{1}{\tau_{e-ph}(\e)} = \dfrac{\delta}{\delta f(\e)}\left(\dfrac{\partial f}{\partial t}\right)_{e-ph}$,
\begin{multline}\label{eq:invtau_ephn}
\dfrac{1}{\tau_{e-ph}(\e)} = \dfrac{D^2}{4\pi\rho (\hbar v_{ph})^4} \sqrt{\dfrac{m_e^\ast}{2\e(1 + C \e)}} \int_0^{\e_{ph}^\text{max}(\e)}(\e_{ph})^2 d\e_{ph} \\ 
\Big[(n_B(\e_{ph},T_{ph}) + f(\e + \e_{ph}))(1 + 2C(\e + \e_{ph})) + \\ 
(n_B(\e_{ph},T_{ph}) + 1 - f(\e - \e_{ph}))(1 + 2C(\e - \e_{ph})) \Big].
\end{multline}
% can be \Gamma_ph!
The $e-ph$ collision rate~\eqref{eq:invtau_ephn} is plotted in Fig.~\ref{fig:dos_invtau}(b). It grows monotonically up to $k(\e)=q_D/2$ ($\e \sim 2.8\ \e_F$), 
%$\e \sim \e_\text{min,D} \sim 2.8\ \e_F$, 
beyond which point its energy-dependence becomes much weaker. 

To gain more insight into the dependence of $\tau_{e-ph}^{-1}$ on material parameters, we simplify Eq.~\eqref{eq:invtau_ephn} by expanding its integrand in a power series in $\e_{ph}$ since the phonon energy is small relative to the electron energy. After integration of the leading-order term of the integrand ($k_B T_{ph} \e_{ph}$) one gets % The integrand of Eq.~\eqref{eq:invtau_ephn} is $\sim (\e_{ph})^2n_b(\e_{ph},T_{ph})$ and thus is $\sim \e_{ph}$ in the leading-order ($n_b(\e_{ph},T_{ph})\sim\dfrac{k_B T_{ph}}{\e_{ph}}$ for small $\e_{ph}$). As a result, after integral over $\e_{ph}$, $\tau_{e-ph}^{-1}$ is proportional to $(\e_{ph}^\text{max})^2$. 
%~\footnote{Although Eq.~\eqref{eq:invtau_ephn_simp} is a good approximation of Eq.~\eqref{eq:invtau_ephn}, it does not have the dip-like feature near $\e_F$ seen in Fig.~\ref{fig:dos_invtau}(b). To resolve this, one needs to incorporate higher-order terms in the expansion of the integrand in Eq.~\eqref{eq:invtau_ephn}.}
\begin{align}\label{eq:invtau_ephn_simp}
%\dfrac{1}{\tau_{e-ph}(\e)}\approx\dfrac{D^2 k_B T_{ph}}{4\pi\rho(\hbar v_{ph})^4}\sqrt{\dfrac{m_e^\ast}{2\e(1 + C\e)}}(1 + 2C\e)\times
%\begin{cases}
%8 m_e^\ast v_{ph}^2 \e(1 + C\e) & \text{for }\e \leq \e_\text{min,D} \\
%\e_D^2&\text{for }\e > \e_\text{min,D}
%\end{cases}
\dfrac{1}{\tau_{e-ph}(\e)}\approx\dfrac{D^2 k_B T_{ph}}{4\pi\rho(\hbar v_{ph})^2}\dfrac{m_e^\ast}{\hbar k(\e)}(1 + 2C\e)\times
\begin{cases}
4 k^2(\e), & \text{for }k(\e) < q_D/2 \\
q_D^2, & \text{for }k(\e) \geq q_D/2
\end{cases},
\end{align}
and shows a descent agreement with the exact expression~\eqref{eq:invtau_ephn}, see Fig.~\ref{fig:dos_invtau}(b). The mismatch between Eq.~\eqref{eq:invtau_ephn} and~\eqref{eq:invtau_ephn_simp}, including the dip-like feature near $\e_F$, can be resolved by incorporating higher-order terms. Eq.~\eqref{eq:invtau_ephn_simp} shows that $\tau_{e-ph}^{-1}$ is proportional to the phonon temperature as in noble metals~\cite{BTl}, and that non-parabolicity increases $\tau_{e-ph}^{-1}$ by a factor of $k(\e_F)(1 + 2 C \e_F)\sim \sqrt{1 + C \e_F}(1 + 2 C \e_F)\sim 2$, similar to the electron density of states, as shown in Fig.~\ref{fig:dos_invtau}(b). 

As a comparison, we also plot the $e-ph$ collision rate as calculated without accounting for momentum conservation. This shows that the $e-ph$ collision rate can be overestimated if momentum conservation is neglected. Indeed, due to the low electron density, the Fermi momentum ($k_F \approx 3.54$ nm\textsuperscript{-1}) is much smaller than the Debye momentum of ITO ($q_D \approx 14.3$ nm\textsuperscript{-1}) so that a substantial amount of the phonons are prohibited from interacting with the electrons due to conservation of momentum. This can be further understood using a phase-space argument, as detailed in Appendix~\ref{app:e-ph_phase_space}. In particular, ignoring momentum conservation causes a $\sim 5$-fold over-estimate of the $e-ph$ collision rate around the Fermi energy (see Fig.~\ref{fig:dos_invtau}(b); this is the most relevant energy regime, see Fig.~\ref{fig:BE_terms}(c) below); consequently, this causes a 30-fold over-estimate of the energy transfer rate between the electrons and phonons (see Section~\ref{subsec:TTM}). This behaviour contrasts the situation in noble metals, for which the Debye momentum (e.g., $q_D \approx 6.85$ nm\textsuperscript{-1} and $q_D \approx 8.14$ nm\textsuperscript{-1} for Au and Ag, respectively) is smaller than their Fermi momentum ($k_F \approx 11.57$ nm\textsuperscript{-1} and $k_F \approx 12$ nm\textsuperscript{-1} for Au and Ag, respectively). For that reason, in noble metals almost all electrons can interact with all phonons such that neglecting momentum conservation is justified. Therefore, somewhat peculiarly, the contradicting effects of the higher Debye energy in ITO (which make $\tau_{e-ph}^{-1}$ 4 times larger, see Eq.~(\ref{eq:invtau_ephn_simp})) on one hand, and the limitations on the allowed $e-ph$ scattering events imposed by the momentum conservation (which make it $\sim 5$ times smaller) on the other hand, make the overall magnitude of the $e-ph$ collision rate in ITO comparable to that in noble metals.

\subsection{The electron-electron ($e-e$) collision term}\label{sub:e-e}
The $e-e$ interaction is responsible for the thermalization of the conduction electrons. The corresponding collision term is derived from the screened Coulomb interaction $U$ whose potential in momentum space is written by %Hamiltonian, namely,
%\begin{align}
%\mathcal{H}_{e-e} = \dfrac{1}{2V}\sum_{{\bf k},{\bf k}_1,{\bf k}_2,{\bf k}_3} U(|{\bf k}-{\bf k}_1+{\bf k}_2-{\bf k}_3|/2)c^\dagger_{{\bf k}_3}c^\dagger_{{\bf k}_2}c_{{\bf k}_1}c_{{\bf k}}\delta_{{\bf k}+{\bf k}_1,{\bf k}_2+{\bf k}_3},
%\end{align}
%where $U(q)$ is the Fourier component of the screened Coulombic interaction
\begin{align}
U({\bf q}) = \dfrac{e^2}{4\pi\varepsilon_0\varepsilon_b }\int d^3r\dfrac{e^{-q_\text{TF} r}}{r} e^{- i {\bf q}\cdot{\bf r}} = \dfrac{e^2}{4\pi\varepsilon_0 \varepsilon_b}\dfrac{1}{q^2 + q_\text{TF}^2},
\end{align}
% define e and epsilon_0...
where ${\bf q}$ is the exchange of momentum between two electrons, and $q_\text{TF}$ is the Thomas-Fermi wave vector which is given by~\cite{delFatti_nonequilib_2000} %{\bf IW/SS - it was above the phonon momentum... is this standard? worth adapting the notations? - This is the notation being used in the literature. It share the same notation as the phonon momentum because both of them are the momentum exchange in the scattering process.}, $q_\text{TF}$ is the Thomas-Fermi wave vector given by~\cite{delFatti_nonequilib_2000} %{\bf the mixed momentum/energy notation is standard? apparently yes}
\begin{align}\label{eq:TF_q_vec}
q^2_{\text{TF}} = -\dfrac{4\pi^3e^2}{\varepsilon_0\varepsilon_b}\int d\e\dfrac{\partial f}{\partial \e}\rho_e(\e),
\end{align}
where $\varepsilon_b = 4$ represents the contribution of interband transitions to the permittivity~\cite{Xian_group_ITO_2019,Sapienza_2022}. Again, following the Fermi's golden rule employed in~\cite{Snoke_solid_state} and accounting for the non-parabolicity of the conduction band, we obtain the population distribution change rate
\begin{multline}\label{eq:dfEdt_ee}
\left(\dfrac{\partial f(\e)}{\partial t}\right)_{e-e} = -\dfrac{3}{16\hbar^2}\dfrac{1}{(2\pi)^3}\left(\dfrac{e^2}{\varepsilon_0\varepsilon_b}\right)^2\left(\dfrac{m_e^\ast}{\hbar^2 q_\text{TF}}\right)^3\dfrac{\hbar}{\sqrt{2 m_e^\ast (\e+C\e^2)}} \int d\e_1d \e_2 d\e_3 \\
(1 + 2 C \e_1)(1 + 2 C \e_2)(1 + 2 C \e_3) \left[\dfrac{qq_\text{TF}}{q^2+q^2_\text{TF}} + \tan^{-1}\left(\dfrac{q}{q_\text{TF}}\right)\right]^{q^\text{max}}_{q^\text{min}} \delta(\e + \e_1 - \e_2 - \e_3)\\
\left[ (1 - f(\e_3))(1 - f(\e_2))f(\e_1)f(\e) - (1 - f(\e))(1 - f(\e_1))f(\e_2)f(\e_3) \right],
\end{multline}
where $q^\text{min} = \max(|k - k_2|,|k_1 - k_3|)$ and $q^\text{max} = \min(k + k_2,k_1 + k_3)$. By taking the functional derivative of Eq.~\eqref{eq:dfEdt_ee} with respect to $f(\e)$, we obtain the collision rate associated with the $e-e$ interaction,
\begin{multline}\label{eq:invtau_ee}
\dfrac{1}{\tau_{e-e}(\e)} = \dfrac{3}{16\hbar^2}\dfrac{1}{(2\pi)^3}\left(\dfrac{e^2}{\varepsilon_0\varepsilon_b}\right)^2\left(\dfrac{m_e^\ast}{\hbar^2 q_\text{TF}}\right)^3\dfrac{\hbar}{\sqrt{2m_e^\ast (\e+C\e^2)}} \int d\e_1d\e_2d\e_3 \\
(1 + 2C\e_1)(1 + 2C\e_2)(1 + 2C\e_3) \left[\dfrac{qq_\text{TF}}{q^2 + q^2_\text{TF}} + \tan^{-1}\left(\dfrac{q}{q_\text{TF}}\right)\right]^{q^\text{max}}_{q^\text{min}} \delta(\e + \e_1 - \e_2 - \e_3) \\
\left[(1 - f(\e_3))(1 - f(\e_2))f(\e_1) + (1 - f(\e_1))f(\e_2) f(\e_3) \right].
\end{multline}
To gain more insight into the non-parabolicity correction to the $e-e$ collision rate, we substitute the electron distribution in the intergand of Eq.~\eqref{eq:invtau_ee} by the Fermi-Dirac distribution function characterized by an electron temperature $T_e$; after some algebra, we find
\begin{align}\label{eq:FLT}
\dfrac{1}{\tau_\text{e-e}(\e)} \approx  K \left\{\left[(\pi k_B T_e)^2 + (\e - \e_F)^2\right] + \dfrac{4}{3}\dfrac{C(\e - \e_F)}{(1 + 2C \e_F)^3}(\pi k_B T_e)^2 + \cdots\right\},
\end{align}
where the prefactor $K$ is given by
\begin{align}\label{eq:K}
K = \dfrac{3}{32\hbar}\dfrac{1}{(2\pi)^3} \left(\dfrac{e^2}{\varepsilon_0 \varepsilon_b}\right)^2 \left(\dfrac{m_e^\ast}{\hbar^2 q_\text{TF}}\right)^3 \dfrac{1}{k_F} \left[\dfrac{2 k_F q_\text{TF}}{4k_F^2 + q^2_\text{TF}} + \tan^{-1}\left(\dfrac{2k_F}{q_\text{TF}}\right)\right](1 + 2 C\e_F)^3.
\end{align}
This expression provides the generalization of Fermi liquid theory~\cite{Quantum-Liquid-Coleman} to Drude materials with non-parabolic bands. 
% $K$ represents the characteristic $e-e$ scattering rate. %{\bf is it extracted a-posteriori by fitting to which experiment? what are the values we have? like in Au/Ag? - $K$ is not a fitting parameter. : No, it is not a fit parameter in ITO, we have used the electron density to determine $q_{TF}$ which determines $K$. }
Compared with the case of $C = 0$, the non-parabolicity overall causes an increase of the density of states such that the $e-e$ thermalization rate increases by a factor of $(1 + 2 C\e_F)^{3/2} (1 + C\e_F)^{-5/4} $~\footnote{From Eq.~\eqref{eq:TF_q_vec}, one can deduce that the Thomas-Fermi wavevector is proportional to the square root of the eDOS at the Fermi-energy~\cite{Ashcroft-Mermin}. Therefore, compared with the case of $C = 0$, $q_{TF}$ is larger by a factor of $(1 + C\e_F)^{1/4}(1 + 2C\e_F)^{1/2}$ and $k_F$ is larger by a factor of $(1 + C\e_F)^{1/2}$. Together with the factor $(1 + 2 C \e_F)^3$ coming from the increase of eDOS, the $e-e$ thermalization rate increases by a factor of $\sim (1 + 2 C\e_F)^{3/2} (1 + C\e_F)^{-5/4}$.} ($\approx 1.5$ for ITO), and adds a small correction term which is linear with $\e-\e_F$~\footnote{Note that the term $(\e - \e_F)$ in Eq.~(\ref{eq:FLT}) should not be replaced by $\hbar \omega$, since it involves all possible electron states, rather than only those excited from the Fermi energy by photon absorption.}, see Fig.~\ref{fig:dos_invtau}(c). Eq.~(\ref{eq:K}) also shows the effect of the low electron density (hence, smaller Fermi energy) in ITO - not only it results in a narrower energy-dependence of $\tau^{-1}_\text{e-e}(\e)$, it also results in smaller Fermi momentum, and more importantly, in weaker screening, and thus, a smaller Thomas-Fermi wavevector~\footnote{The Thomas-Fermi wavevector is proportional to the square root of the eDOS at the Fermi-energy~\cite{Ashcroft-Mermin}, thus, it decreases with electron density.}. This means that %Because, smaller thomas-fermi wave-vector means slower (exponential) decay in Yukawa potential $V(\mathbf{r}) = \frac{e}{r} e^{-q_{TF} r}$ leading to longer range interaction and faster scattering},
$e-e$ interactions are stronger in ITO, leading to a $\sim 10$ times faster $e-e$ collision rate than Au. %{\bf so we are closer to the Wilson and Coh's condition... maybe the condition should be amended to refer to the microscopic rates?}

%characteristic $e-e$ scattering rate of ITO $K\approx0.04~\text{eV}^{-2}\cdot\text{fs}^{-1}$ is $\sim 10$ times larger than that of Au. 

%\begin{figure}[h]
%\centering
%\includegraphics[width=0.4\textwidth]{figures/eta_ee_C.eps}
%\caption{(Color online) The relaxation rate associated with the $e-e$ interaction with (blue solid line, $C\neq 0$) and without (blue dashed line, $C=0$) non-parabolicity.}
%\label{fig:eta_ee_C}
%\end{figure}
% BE without RTA
% doing this is not as numerically challenging as in Au, because of the smaller band and higher Debye temperature that enables coarser energy steps.

% for much lower densities the trend should change to a decrease and eventually independence on density at intrinsic SC densities. Snoke never calculated this transition, although it is non-trivial/non-intuitive and relevant for highly doped SCs.

% problems in Boyd~\cite{Boyd_Nat_Phot_2018} - formulate for $e-e$ collision rate wrong - uses photon instead of electron energy + use it in a macroscopic setting rather than a microscopic one. uses mathiessen's rule... .

\section{Results}
In the example below, we solve the BE~(\ref{eq:f_neq_dynamics}) in order to obtain the electronic (i.e., solid-state physics) response of bulk ITO systems subject to (modestly high level) pulsed electric field illumination $I_\text{inc} = I_0 e^{- 4 \ln(2) (t/\tau_p)^2}$, a wavelength of $1300$ nm, duration of $\tau_p = 220$ fs and peak intensity of $I_0 = 2.5$ GW/cm$^2$. However, as for the uncertainty on the various material parameters, the results below remain generic also for other parameter values.

% so more states due to non-parabolicity

\subsection{electron dynamics}\label{subsec:e-dynamics}
\begin{figure}[h]
\centering
\includegraphics[width=0.95\textwidth]{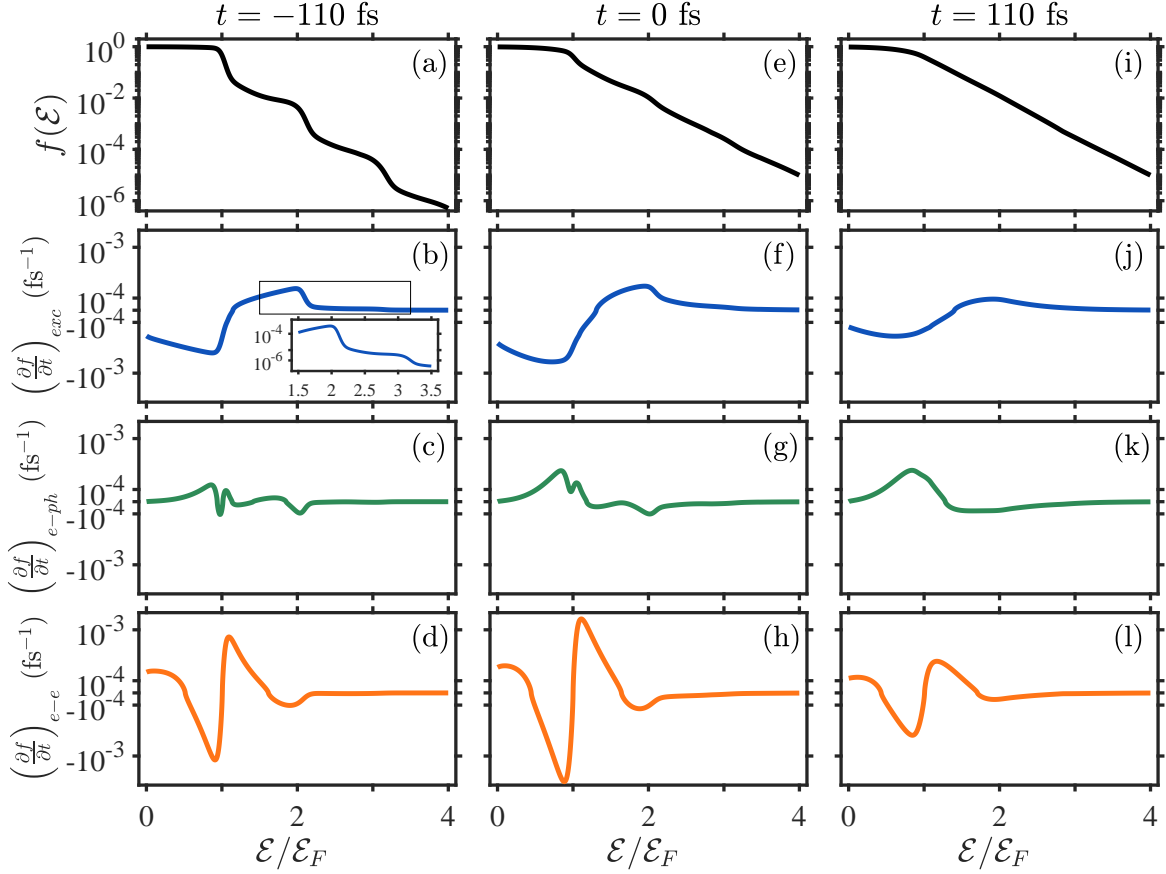}
\caption{(Color online) %(a) Photo-excitation (blue) (b) electron-phonon interaction (green), and (c) electron-electron interaction (orange) terms at the FWHM \textit{before} the peak of the pulse (dashed; $t = -110$ fs), \textit{at the centre} of the pulse (solid: $t =0$ fs) and at the FWHM \textit{after} the peak of the pulse (dot-dashed line; $t = +110$ fs) as a function of $\mathcal{E}/\mathcal{E}_F$.
(a), (e) and (i) Electron distribution (black lines), (b), (f) and (j) photo-excitation (blue), (c), (g) and (k) electron-phonon interaction (green), and (d), (h) and (l) electron-electron interaction (orange) terms following illumination of ITO by a short pulse. The left column shows the various terms at the FWHM \textit{before} the peak of the pulse ($t = - \tau_p/2 = - 110$ fs), the middle column \textit{at the centre} of the pulse ($t = 0$ fs), and the right column at the FWHM \textit{after} the peak of the pulse ($t = + 110$ fs) as a function of $\mathcal{E}/\mathcal{E}_F$ and pulse peak intensity $I_0 = 2.5$ GW/cm$^2$. The inset in (b) is the zoom-in of the region between $\e = 1.5$ eV and $\e = 3.5$ eV; it shows more clearly the origin of the multiple step structure seen in (a). %showing the reminiscent of the  non-sequential two-photon absorption due to the creation of the non-thermal shoulder between $\e_F$ and $\e_F+\hbar\omega$ in $f(\e)$. This is responsible for creation of the non-thermal shoulder between $\e_F+\hbar\omega$ and $\e_F+2\hbar\omega$ in $f(\e)$. 
All terms have the same order of magnitude because of the pulsed nature of the illumination. The $y$-axis is in log-scale for (a), (e) and (i) and in symmetric log-scale for all other subplots. 
}
\label{fig:BE_terms}
\end{figure}

Fig.~\ref{fig:BE_terms} plots the electron distributions and the different terms corresponding to Boltzmann equation (BE) \eqref{eq:f_neq_dynamics} at three different time slices, $t = - \tau_p/2 = -110$ fs, $t = 0$ fs, and $t = -110$ fs corresponding to full width at the half maximum (FWHM) before the peak, at the peak and FWHM after the peak of the pulse. 

Fig.~\ref{fig:BE_terms}(a) shows the initial electron distribution $f(\e)$ as a function of normalized energy $\e/\e_F$. The deviation from thermal equilibrium is clearly visible via the characteristic shoulders above Fermi energy, corresponding to one, two etc. consecutive photon absorption events; indeed, those were observed previously for noble metals, e.g., in~\cite{non_eq_model_Rethfeld,Italians_hot_es,Sivan-Dubi-PL_I}. They originate from the structure of the excitation term (Fig.~\ref{fig:BE_terms}(b))%~\footnote{In particular, the inset of Fig.~\ref{fig:BE_terms}(b) shows the second $\hbar \omega$-wide region of positive change rate of population. This is reminiscent of the non-sequential two-photon absorption due to the non-thermal shoulder between $\e_F$ and $\e_F + \hbar\omega$ in $f(\e)$, and is also responsible for the creation of the second non-thermal shoulder between $\e_F + \hbar\omega$ and $\e_F + 2\hbar\omega$ in $f(\e)$, see Fig.~\ref{fig:BE_terms}(a). }
. However, these shoulders gradually smooth out due to (the weak) $e-ph$ collisions, and (the much stronger) $e-e$ collisions, see Fig.~\ref{fig:BE_terms}(c) and~(d), respectively. Indeed, at later times the distribution rapidly approaches a thermal distribution (see Fig.~\ref{fig:BE_terms}(e) and~(i)). 

The corresponding evolution of the various terms is seen in the additional subplots of Fig.~\ref{fig:BE_terms}. It is interesting to note the differences with respect to the corresponding dynamics in noble metals. Specifically, due to the low Fermi energy, there is only a single $\hbar \omega$-wide region of negative rate of change of population $\displaystyle \left(\dfrac{df(\e)}{dt}\right)_{exc}$ due to photon absorption below the Fermi energy but a corresponding multiple-shoulder structure of positive rate above the Fermi energy; the various shoulders are energy-dependent due to the relatively strong energy dependence of the eDOS (see Fig.~\ref{fig:dos_invtau}(a)). Moreover, near the band minimum, photon absorption is weaker due the vanishing electron density of states. This leads to a sudden cut-off of $\partial f/\partial t|_{exc}$~\eqref{eq:dfEdt_exc} near the band minimum. 

% Fig.~\ref{fig:BE_terms}(e) and (f) show that after the pulse gets completely absorbed substantial smearing of Eq.~\eqref{eq:dfEdt_exc} happens due to smearing out of the electron distribution $f(\e)$ (which is a result of thermalization). %{\bf the other way around?}.

The structure of the electron-phonon ($e-ph$) term is significantly different compared to its structure in noble metals, see Fig.~\ref{fig:BE_terms}(c), (g) and (k) corresponding to Eq.~\eqref{eq:dfEdt_eph}. The origin of this difference is the importance of momentum conservation along with the number conserving nature of the $e-ph$ interaction which limit the possible scattering processes, see details in Appendix~\ref{app:dfdt_eph_shape}. Also notable is the increase in magnitude of the $e-ph$ term in time; this is related to the rise of the overall electron energy and the rapid increase in the number of low excess energy electrons; this effect was already shown in~\cite{non_eq_model_Lagendijk} to lead to an acceleration in the $e-ph$ rate, which is not captured by the RTA.

%However, due to thermalization, the shape of $\displaystyle \left(\dfrac{df(\e)}{dt}\right)_{e-ph}$ takes the rather well-known shape \cite{Dubi-Sivan} corresponding to positive rate below, see Fig.~\ref{fig:BE_terms}(i) {\bf not really...?} At t = 110 fs, although the electron distribution is nearly thermalized, $\displaystyle \left(\dfrac{df(\e)}{dt}\right)_{e-ph}$ exhibits a different shape from the well-known shape shown in~\cite{Dubi-Sivan}. This is again due to momentum conservation and number conservation in the $e-ph$ process ...see Appendix

% At the peak of the pulse amplitude (Fig.~\ref{fig:BE_terms}(b)), these shoulders already mostly disappear due to the rapid thermalization induced by the fast electron-electron ($e-e$) scattering. Fig.~\ref{fig:BE_terms}(c) indicates that by the time the pulse gets absorbed the electron gas has thermalized almost completely.

% Fig.~\ref{fig:BE_terms} (d)-(f) show the photo-excitation term of the BE~\eqref{eq:dfEdt_exc} which is responsible for the shoulder structure in $f(\e)$. 

Finally, Fig.~\ref{fig:BE_terms} (d), (h) and (l) show the rate of change of $f(\e)$ due to electron-electron ($e-e$) interactions corresponding to Eq.~\eqref{eq:dfEdt_ee}. Its shape is similar to that in noble metals (cf~\cite{Dubi-Sivan}), except near the band minimum and for $\e \approx \e_F + \hbar \omega$. This is due to the energy and number conserving nature of the $e-e$ interaction. Near the band minimum, the positive rate of change of population is enhanced due to the vanishing eDOS.

\subsection{Coarse-grained dynamics - the Two Temperature Model (TTM)}\label{subsec:TTM}

Quite frequently, it is sufficient to consider the macroscopic dynamics of the energies of the electron and phonon subsystems, respectively. This can be achieved first by integrating over the product of the various terms in the BE~(\ref{eq:f_neq_dynamics}) with the electron energy and eDOS (see, e.g.,~\cite{delFatti_nonequilib_2000,Dubi-Sivan}), which provides the resulting total energy of the electron subsystem. Then, a dynamic equation for the phonon energy can be written down based on the total rate of energy transfer from the electron subsystem. In order to make the resulting equations more meaningful, it is customary to rewrite the resulting energies as the product of the respective heat capacities ($C_e$ and $C_{ph}$) and electron and phonon temperatures. While the latter is well-defined, it is well known that the notion of electron temperature cannot be defined clearly in the initial stages of the dynamics~\cite{non_eq_model_Lagendijk,Italians_hot_es,Stoll_review}. In this context, it is customary to extract an {\em effective} electron temperature by calculating the electron temperature $T_e$ for which the total energy of a Fermi Dirac (i.e., thermal) distribution $f^T$~(\ref{eq:f_eq}) is the same as that of the true non-thermal distribution $f(\e)$~(\ref{eq:f_neq_dynamics}) (see, e.g.,~\cite{Italians_hot_es,Stoll_review,GdA_hot_es}), namely, 
\begin{equation}\label{eq:Te_extract}
\mathcal{U}_e \equiv \int \e \rho_e(\e) f(\e) d\e = \int \e \rho_e(\e) f^T(\e,T_e) d\e.
\end{equation}
The resulting {\em effective} electron temperature emerges to be the well-defined electron temperature once the distribution thermalizes~\cite{Stoll_review}. Since $e-e$ interactions conserve the energy of the electron subsystem, the integrated BE emerge to be
\begin{align}\label{eq:dUedt}
C_e\dfrac{dT_e}{dt} &= P_{abs} - G_{e-ph} (T_e - T_{ph}), 
\end{align}
while the equation for the phonon temperature is
\begin{align}\label{eq:dUphdt}
C_{ph}\dfrac{dT_{ph}}{dt} &= \ \ G_{e-ph} (T_e - T_{ph}).
\end{align}
Note that since we are interested in the ultrafast dynamics, we ignore heat transfer to the environment (assumed to be at the environment temperature), as this process occurs on a much longer time scale. Equations~(\ref{eq:dUedt})-(\ref{eq:dUphdt}) constitute the so-called  ``Two Temperature Model''; here, $C_e(T_e)$ and $C_{ph}$ are the electron and phonon heat capacities of ITO, and $G_{e-ph}$ is the electron-phonon coupling. The advantage of such a coarse-grained model is considerable, as it is significantly simpler to solve compared with the BE, and serves as the basis for temperature-based permittivity models. 

The TTM parameters are usually hard to measure directly. For example, the phonon heat capacity was measured to be $C_{ph} = 2.54 \times 10^6$ J$\cdot$m$^{-3} \cdot $K$^{-1}$~\cite{ITO_Cph} %{\bf IW/SS - is this a measured value? can it be computed self-consistently from our model? - the $T_{ph}$-dependent $C_{ph}$ can be evaluated using the Debye model. It shows that $C_{ph}$ increases by $\sim20\%$ when $T_{ph}$ increases from 300 K to 700 K. We can add this to our model and determine $C_{ph}$ self-consistently. But this should mot make a big difference.} % assuming it does not change much with $T$ because of the modest T rise
and similar values arise from a direct calculation. More frequently, the TTM parameters are calculated theoretically using thermal distributions at the effective electron temperature. For noble metals, relatively simple expressions for the various emerging parameters were obtained (e.g.,~\cite{Two_temp_model,Stoll_review}); in particular, both electron heat capacity and $e-ph$ coupling can be written as linear functions of the temperatures. These parameters, however, are significantly more complicated in LEDD materials compared to noble metals because of the non-parabolicity, and especially because of the much smaller electron density (hence, smaller Fermi energy $\e_F$). In principle, the TTM parameters can be evaluated through integral expressions (see below). However, in what follows we also provide approximate analytic expression for these parameters; these expressions are suitable for any value of intrinsic parameters of ITO (or other LEDD materials), which indeed vary due to fabrication/doping conditions (see e.g.,~\cite{Guru_Boltasseva_alternatives,Feigenbaum-Atwater-doping,Liu_ITO_2014}). 

\begin{figure}[h]
\centering
\includegraphics[width=1.0\textwidth]{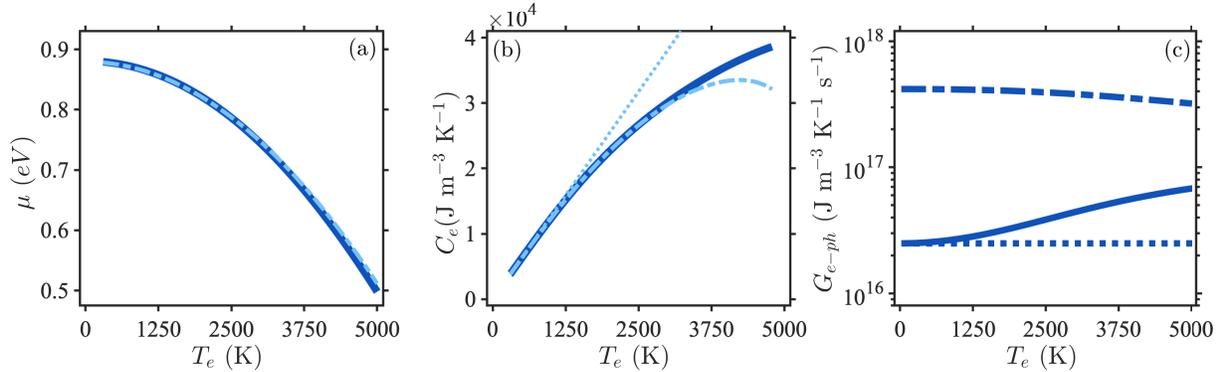}
\caption{(Color online) (a) chemical potential $\mu$ and (b) electron heat capacity $C_e$ as a function of electron temperature $T_e$. The blue solid and dot-dashed lines correspond to the numerical values and analytical forms (Eqs.~\eqref{eq:num_conserv_mue} for $\mu$ and \eqref{eq:C_e_analytical} for $C_e$), respectively. The dotted line in (b) corresponds to the linear approximation of $C_e$ indicating the deviation from the linear dependence above $T_e \sim 1300$ K. (c) The $e-ph$ coupling coefficient $G_{e-ph}$ as a function of the electron temperature~\eqref{eq:G_eph_int} (blue solid line) and its zero temperature limit~\eqref{eq:G_eph_int_0K} (blue dotted line). The blue dashed line represents the case without accounting for momentum conservation~\eqref{eq:G_eph_no_mc}. }
\label{fig:Geph_ce_mu}
\end{figure}

One effect of the smaller $\e_F$ is that the dependence of the chemical potential on temperature is not negligible as in noble metals~\footnote{This is reminiscent of what happens in semiconductors~\cite{Ashcroft-Mermin,SzeCh1,Sarkar-Un-Sivan-Dubi-NESS-SC}. }. To see this, we employ the Sommerfeld expansion~\cite{Sommerfeld_1928,Ashcroft-Mermin} to express the total energy of the electron subsystem $\mathcal{U}_e$ as a Taylor expansion in powers of $k_B T_e$, assuming purely thermal electron distributions. In standard textbooks, e.g.~\cite{Ashcroft-Mermin}, the expansion is usually kept up to the second-order only. However, since the Fermi energy of ITO is much lower than that of metals and since the incoming illumination intensity is strong such that the electron temperature might become non-negligible with respect to the Fermi temperature (e.g., $\sim 10,000~\text{K}$ in ITO), one needs to keep the expansion at least up to the fourth-order, namely, 
\begin{multline}\label{eq:eng_conserv_Te}
\mathcal{U}_e \approx \int_0^{\mu(T_e)}\e\rho_e(\e)d\e + \dfrac{\pi^2}{6}(k_B T_e)^2\left.\dfrac{d\left(\e\rho_e(\e)\right)}{d\e}\right|_{\e = \mu(T_e)} + \dfrac{7\pi^4}{360}(k_B T_e)^4 \left.\dfrac{d^3\left(\e\rho_e(\e)\right)}{d\e^3}\right|_{\e = \mu(T_e)}.
\end{multline}
Then, the temperature-dependent chemical potential $\mu(T_e)$ is determined using number conservation, $\displaystyle \int_0^\infty \rho_e(\e) f^T(\e,\mu(T_e),T_e) d\e = \displaystyle \int_0^{\e_F} \rho_e(\e) d\e $, i.e., 
\begin{align}\label{eq:num_conserv_mue}
%\left[\e_F(1 + C\e_F)\right]^{2/3}
%\approx\left[\mu(T_e)(1 + C\mu(T_e))\right]^{2/3} \\
% + \dfrac{\pi^2}{4}(k_BT_e)^2 \dfrac{1 + 8C\mu(T_e)(1 + C\mu(T_e))}{2\sqrt{\mu(T_e)(1 + C\mu(T_e))}}
%+ \dfrac{7\pi^4}{640} \dfrac{(k_BT_e)^4}{\left[\mu(T_e)(1 + C\mu(T_e))\right]^{5/2}}.
\mu(T_e) \approx \e_F - \dfrac{(\pi k_B T_e)^2}{6} \dfrac{1}{\rho_e(\e_F)} \left.\dfrac{d\rho_e(\e)}{d\e}\right|_{\e_F},
\end{align}
where $\displaystyle \int_0^{\e_F} \rho_e(\e) d\e$ is the number of electrons at zero-temperature (which is indeed the same as that at 300K). This expression is plotted in Fig.~\ref{fig:Geph_ce_mu}(a) vs. the exact numerical solution.

Similarly to $\mu$, the exact integral definition of the electron heat capacity can be approximated as
\begin{multline}\label{eq:C_e_analytical}
C_e(T_e) \equiv \frac{d \mathcal{U}_e}{d T_e} = \frac{d}{d T_e}\left[\int \e \rho_e(\e) f^T(\e,T_e) d\e \right] \\
\approx \dfrac{\pi^2 k_{B}^{2}}{3} T_e \left[\rho_e (\mu) + \mu \rho_e^{(1)}(\mu) \right] + \dfrac{7\pi^4 k_B^{4}}{90} T_e^3 \left[3 \rho_e^{(2)} (\mu) + \mu \rho_e^{(3)}(\mu) \right] \\ + \dfrac{d \mu}{d T_e} \left( \mu \rho_e(\mu) + \dfrac{(\pi k_B T_e)^2}{6} \left[2\rho_e^{(1)}(\mu) + \mu \rho_e^{(2)}(\mu) \right] + \dfrac{7 (\pi k_B T_e)^4}{360} \left[2\rho_e^{(3)}(\mu) + \mu \rho_e^{(4)}(\mu) \right] \right),
%\int_0^{\mu(T_e)}\e\rho_e(\e)d\e + \dfrac{\pi^2}{6}(k_B T_e)^2\left.\dfrac{d\left(\e\rho_e(\e)\right)}{d\e}\right|_{\e=\mu(T_e)} + \dfrac{7\pi^4}{360}(k_B T_e)^4 \left.\dfrac{d^3\left(\e\rho_e(\e)\right)}{d\e^3}\right|_{\e = \mu(T_e)},
\end{multline}
where $\displaystyle \rho_e^{(n)}(\mu) = \left. \dfrac{d^n \rho_e (\e)}{d \e^n} \right|_\mu$ and $\mu(T_e)$ is given by Eq.~\eqref{eq:num_conserv_mue}. As shown in Fig.~\ref{fig:Geph_ce_mu}(b), up to $T_e \sim 1300$K, the electron heat capacity scales linearly with the electron temperature (viz., $C_e \approx \gamma_e T_e$ with $\gamma_e = \pi^2 k_B^2 \left[\rho_e (\mu) + \mu \rho_e^{(1)}(\mu) \right]/3$~\footnote{Due to non-parabolicity the dependence of $\gamma_e$ on electron density becomes is not straightforward. In contrast, in its absence, $\gamma_e$ reduces to the familiar expression $\gamma_e = \pi^2 n_e k_B^2 / 2 \e_F$.}). Note that the value of $\gamma_e$ for ITO ($12.7~\text{J}\cdot\text{m}^{-3} \cdot\text{K}^{-2}$) is much smaller than for noble metals, e.g., $\gamma_e = 67.6~\text{J} \cdot \text{m}^{-3} \cdot \text{K}^{-2}$ for gold (Au)~\cite{G_e_ph_Zhigilei}. This smaller value of $\gamma_e$ in ITO is associated with the lower electron density; indeed, compared to noble metals, the eDOS is evaluated at the much lower chemical potential, giving rise to a smaller value for $\gamma_e$. %Beyond $T_e \sim 3500$ K the validity of the analytical form \eqref{eq:C_e_analytical} (and therefore the Sommerfeld expression) is ruined due to $k_B T_e$ becoming larger than $\mu(T_e)$.
Furthermore, the cubic dependence of $C_e$ (emerging from the 4th-order term in the Sommerfeld expansion of the electron energy) provides decent accuracy only up to $T_e \sim 3500$ K. % but proper tuning of "C_e" in earlier work might have given the correct dynamics..
In particular, in this regime of electron temperatures, $C)e$ experiences a sublinear growth due to the decrease of $\mu$ with temperature.

Lastly, the $e-ph$ coupling coefficient $G_{e-ph}$ can be obtained by evaluating the energy transferred from the electron to the phonon subsystem, i.e., $\left(\dfrac{\partial \mathcal{U}_e}{\partial t}\right)_{e-ph} = \displaystyle\int \e\rho_e(\e)\left(\dfrac{\partial f}{\partial t}\right)_{e-ph} d\e$, where $\left(\dfrac{\partial f}{\partial t}\right)_{e-ph}$ is given by Eq.~\eqref{eq:dfEdt_eph}. As for $C_e$, we substitute the Fermi-Dirac distribution at the effective electron temperature for $f$ and expand the integrand in a power series in $\e_{ph}$. In similarity to noble metals, we find that the leading order term is proportional to the difference between the electron and phonon temperatures, i.e., $\left(\dfrac{\partial \mathcal{U}_e}{\partial t}\right)_{e-ph} \approx - G_{e-ph}(T_e - T_{ph})$, where the $e-ph$ coupling coefficient is given by
\begin{multline}\label{eq:G_eph_int}
G_{e-ph}(T_e) = \dfrac{D^2 m_e^\ast k_B}{4(\pi\hbar)^3 \rho(\hbar v_{ph})^4}  \int_0^\infty d\e
\int_0^{\e_D}(\e_{ph})^2 d\e_{ph} \\
\left[(1 + 2 C \e) \sech^2\left(\dfrac{\e - \mu}{2k_B T_e}\right)\tanh\left(\dfrac{\e - \mu}{2 k_B T_e}\right) - 4 C k_B T_e\sech^2\left(\dfrac{\e - \mu}{2 k_B T_e}\right)\right] \\
- \dfrac{D^2 m_e^\ast k_B}{4(\pi\hbar)^3\rho(\hbar v_{ph})^4}\Bigg\{
\int_0^\infty d\e \int_0^{\e^\text{max}_{ph}}(\e_{ph})^2 d\e_{ph}\\
\bigg[(1 + 2C\e)\sech^2\left(\dfrac{\e - \mu}{2 k_B T_e}\right) \tanh\left(\dfrac{\e - \mu}{2 k_B T_e}\right) - 4 C k_B T_e \sech^2\left(\dfrac{\e - \mu}{2 k_B T_e}\right)\bigg] \\
- \int_0^{\e_\text{min,D}} d\e
\left[\dfrac{8(m_e^\ast v_{ph}^2)^2}{k_B T_e}\e^2 (1+C\e)(1 + 2C\e)^3 \sech^2\left(\dfrac{\e - \mu}{2k_B T_e}\right)\right]\Bigg\},
\end{multline}
where $\e_{ph}^\text{max}$ is given by Eq.~\eqref{eq:eng_phn_max}. % For $C = 0$, $\e_\text{min} = \displaystyle\lim_{C\rightarrow0}\dfrac{1}{2C}\left(\sqrt{1 + \dfrac{C(\e_{ph})^2}{2m_e^\ast v_{ph}^2}} - 1\right) = \dfrac{(\e_{ph})^2}{8m_e^\ast v_{ph}^2}$}. 

Similar to the other TTM parameters, the $e-ph$ coupling coefficient $G_{e-ph}$ attains a different and more complex form compared to noble metals. Indeed, we plot the electron temperature dependence of $G_{e-ph}$~\eqref{eq:G_eph_int} in Fig.~\ref{fig:Geph_ce_mu}(c). First, we find that the magnitude of $G_{e-ph}$ of ITO ($\sim 3\times10^{16}$ Jm\textsuperscript{-3}K\textsuperscript{-1}s\textsuperscript{-1}) is similar to that of Au ($\sim 2.5 \times 10^{16}$ Jm\textsuperscript{-3}K\textsuperscript{-1}s\textsuperscript{-1}~\cite{PB_Allen_e_ph_scattering,delFatti_nonequilib_2000}~\footnote{For Au, Eq.~\eqref{eq:G_eph_no_mc} becomes $\displaystyle G_{e-ph}^{(Au)} = \dfrac{D^2 m_e k_B \e_D^4}{16(\pi\hbar)^3 \rho(\hbar v_{ph})^4}$; with the known parameters for Au~\cite{delFatti_nonequilib_2000}, (viz., deformation potential $D = 19.3$ eV, Debye energy $\e_D = 0.014$ eV, mass density $\rho = 19.3 \times 10^3~\text{Kg}/\text{m}^3$ and phonon velocity $v_{ph} = 3240 \text{m/s}$) one obtains the value above.}) although the Debye energy of ITO is 4 times larger than that of Au. The reason is that conservation of momentum prohibits a large portion of the phonons from interacting with the electrons, similar to the $e-ph$ collision rate in Section~\ref{sub:e-ph}. %that the Debye momentum $\hbar q_D = \e_D/v_{ph}$ of ITO is 4 times larger than its Fermi momentum, such that a large portion of the phonons are prohibited from interacting with the electrons due to conservation of momentum, see Appendix \ref{app:e-ph_phase_space}. 
Secondly, Fig.~\ref{fig:Geph_ce_mu}(c) shows that quite different from noble metals, $G_{e-ph}$ of ITO increases with the electron temperature. This is again a direct result of the momentum conservation constraint. In general, the momentum conservation in $e-ph$ collision causes that electrons with momentum $\hbar k$ can only absorb/emit phonons with momentum ranging from 0 to $2\hbar k$ (see details in Appendix~\ref{app:e-ph_phase_space}). In ITO, the momentum of many of the electrons is much smaller than the Debye momentum (due to $k_F = q_D/4$), so the maximal momentum of a phonon which can be absorbed/emitted by an electron increases with the momentum of the electron. Therefore, when the electron temperature increases, more electrons occupy higher energy states; these higher energy electrons can then interact with higher energy phonons, leading to faster transfer of energy from the electrons to the phonon subsystem and, thus, to an increase in $G_{e-ph}$~\footnote{The same reasoning explains why an increase in the phonon temperature does not have a significant effect, see Eq.~\eqref{eq:G_eph_int}.}. In contrast, in noble metals the momentum of electrons is much larger than the Debye momentum so that the maximal energy of a phonon which can be absorbed/emitted by electrons is limited by the Debye momentum instead of the electron momentum, so that $G_{e-ph}$ is independent of the electron temperature~\cite{delFatti_nonequilib_2000,vallee_nonequilib_2004,vallee_nonequilib_2003,Dubi-Sivan-APL-Perspective,Dubi-Sivan}.

To gain more insight into $G_{e-ph}$, we analyze Eq.~\eqref{eq:G_eph_int} in some simple limits. First, we consider $e-ph$ coupling coefficient without accounting for momentum conservation. %We start by 
This corresponds to the first term (the first two lines) in Eq.~\eqref{eq:G_eph_int}, and it %. This term corresponds to the $e-ph$ coupling coefficient without accounting for momentum conservation 
can be simplified to be (see Eq.~\eqref{eq:Gephn0_int})
\begin{align}\label{eq:G_eph_no_mc}
G_{e-ph}(T_e) = \dfrac{D^2 m_e^{\ast 2} k_B \e_D^4}{16(\pi\hbar)^3 \rho(\hbar v_{ph})^4} (1 + 2 C \mu(T_e))^2 \qquad \text{(without momentum conservation).}
\end{align}
This approximation is widely used in modeling of noble metals~\cite{delFatti_nonequilib_2000,vallee_nonequilib_2004,vallee_nonequilib_2003,Dubi-Sivan-APL-Perspective,Dubi-Sivan}, however, it is poor for ITO. Indeed, the ratio between Eq.~\eqref{eq:G_eph_int} and Eq.~\eqref{eq:G_eph_no_mc} shows that $G_{e-ph}$ is smaller than that of the case without accounting for momentum conservation by a factor of $\left(\dfrac{\e_D}{2\hbar v_{ph} k_F}\right)^4$ ($\sim 30$), see Fig.~\ref{fig:Geph_ce_mu}(c). %This happens because when the electron temperature becomes higher, more electrons occupy higher energy states, so that electrons can interact with more phonons ($\e_{ph}_\text{max}(\e)$ increases with electron energy, see Eq.~\eqref{eq:eng_phn_max}) and thus transfer their energy to phonons faster. % Finally, $G_{e-ph}$ is larger by a factor of $(1 + 2 C \e_F)^2$ ($\approx 3$ for ITO) compared with the case without non-parabolicity.
Furthermore, due to the non-parabolicity, the $e-ph$ coupling coefficient~\eqref{eq:G_eph_no_mc} is larger by an overall factor of $(1 + 2 C \mu)^2$ ($\sim 3$ for ITO) and shows an (incorrect) decrease with the electron temperature (due to the decrease of chemical potential $\mu$ with the electron temperature $T_e$), see Fig.~\ref{fig:Geph_ce_mu}(c). %The remaining term (the last three lines) in Eq.~\eqref{eq:G_eph_int} subtracts the contribution from the phonons that are restricted by the conservation of momentum. This contribution is usually negligible for noble metals because $\e_\text{min,D}$ (see Eq.~\eqref{eq:eng_phn_max}) is much smaller than their Fermi energy, but is substantial for materials such as ITO (see Appendix~\ref{app:e-ph_phase_space}). %The integral in Eq.~\eqref{eq:G_eph_int} can be evaluated analytically (see Appendix~\ref{app:G_eph_analy_expres}), and 
For a more accurate approximation, we consider the $e-ph$ coupling coefficient with momentum conservation in the zero temperature limit. In this case, Eq.~\eqref{eq:G_eph_int} becomes (see Appendix~\ref{app:G_eph_analy_expres})
\begin{align}\label{eq:G_eph_int_0K}
G_{e-ph}(T_e \rightarrow0\ \text{K}) = \dfrac{D^2 m_e^{\ast2} k_B k_F^4}{\pi^3\rho\hbar^3 }(1 + 2 C \e_F)^2\quad  (k_F<q_D/2\text{ and momentum conserved)}.
\end{align}

\begin{figure}[h]
\centering
\includegraphics[width=0.8\textwidth]{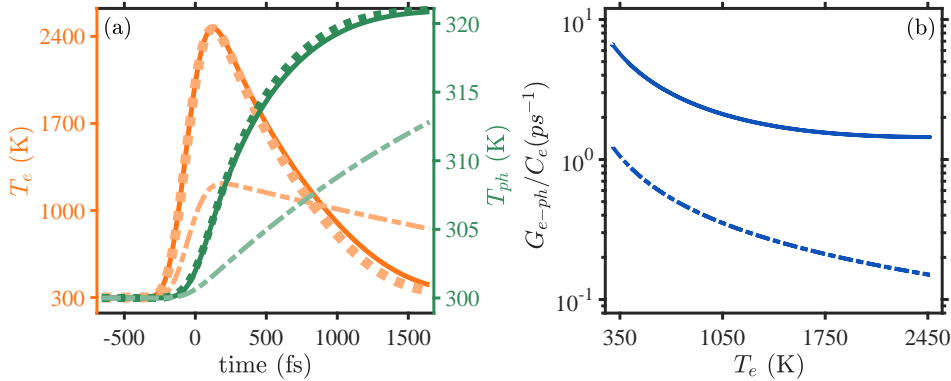}
\caption{(a) (Color online) Effective electron temperatures (orange lines) and phonon temperatures (green lines) as a function of time for the pulse parameters specified in the text. Solid and dotted lines correspond to temperatures obtained from numerical calculation using the BE (using Eq.~\eqref{eq:Te_extract} and \eqref{eq:dUphdt}) % the equation for determining $T_{ph}$ is basically the same in both BE formalism and TTM. From BE we calculate $\left(\dfrac{\partial \mathcal{U}_e}{\partial t}\right)_{e-ph} = \displaystyle\int \e\rho_e(\e)\left(\dfrac{\partial f}{\partial t}\right)_{e-ph} d\e $ directly instead of $G_{e-ph} (T_e - T_{ph})$. In TTM we use \eqref{eq:dUphdt}.}
and the solution of the two-temperature model (TTM) (Eqs.~\eqref{eq:dUedt}-\eqref{eq:dUphdt}), respectively. Dot-dashed lines correspond to the temperature dynamics of gold (Au) obtained from the TTM simulation with $C_e = \gamma_e T_e$ (with $\gamma_e = 67.6~\text{J}\cdot\text{m}^{-3} \cdot\text{K}^{-2}$~\cite{G_e_ph_Zhigilei}) and electron-phonon coupling $G_{e-ph} = 2.5 \times 10^{16}~\text{J}\cdot \text{m}^{–3} \cdot\text{K}^{–1} \cdot \text{s}^{–1}$~\cite{PB_Allen_e_ph_scattering,delFatti_nonequilib_2000}. (b) The ratio $G_{e-ph}/C_e$ for ITO (solid line) and Au (dot-dashed line). }\label{fig:Te_Tphn_dyn}
\end{figure}

Having determined the various parameters appearing in the TTM equations, we can solve them, plot the resulting (effective) electron as well as the phonon temperature dynamics in Fig.~\ref{fig:Te_Tphn_dyn}, and compare it to the dynamics in noble metals, e.g., gold (Au), for the same heat source~(\ref{eq:P_abs}). Overall, the dynamics in these two systems is qualitatively similar, namely, the electron temperature grows on a few 100 fs timescale (dictated by the pulse duration), and then decays due to $e-ph$ interactions. However, the total electron heating in ITO is much higher than in Au. The reason for that is the difference in the values of the corresponding heat capacities; indeed, at low temperatures the electron heat capacity is linear with the electron temperature (i.e., $C_e \sim \gamma_e T_e$), and $\gamma_e$ for ITO is about 5 times smaller than in Au (as already mentioned above, $\gamma_e = 67.6~\text{J}\cdot\text{m}^{-3} \cdot\text{K}^{-2}$ in Au~\cite{G_e_ph_Zhigilei} and $\gamma_e = 12.7~\text{J}\cdot\text{m}^{-3} \cdot\text{K}^{-2}$ obtained for ITO). As a rough estimate, one can ignore $e-ph$ heat transfer in the initial stages of the dynamics, so that the temperature rise can be easily shown to scale as $\sqrt{\mathcal{U}_\text{abs} / \gamma_e}$ (see, e.g.,~\cite{Lupton_transient_metal_PL}). The ratio of the maximal temperature rise in ITO and Au ($\sim 2100$ K and $\sim 900$ K, respectively) is indeed given roughly by $\sqrt{67.6/12.7} \sim 2.3$.

Another notable difference is that the decrease rate of the (effective) electron temperature (and correspondingly, the rise time of the phonon temperature) is faster in ITO compared to Au, see Fig.~\ref{fig:Te_Tphn_dyn}(a). To leading order, these rates are determined by the ratio $G_{e-ph}/C_e$; since $C_e$ is lower in ITO, but $G_{e-ph}$ is comparable in ITO and Au, the rates in ITO are higher, see Fig.~\ref{fig:Te_Tphn_dyn}(b). Nevertheless, since the phonon heat capacity and the heat absorption is similar in both ITO and Au, the eventual phonon temperature reached is similar in the two systems (not shown).

%\begin{figure}[h]
%\centering
%\includegraphics[width=0.45\textwidth]{figures/ce_compare.eps}
%\caption{(Color online) merged with Fig. 6. Chemical potential $\mu$ in eV (blue lines) and electronic specific heat $C_e/k_B$ in $\text{mol}^{-1} \text{K}^{-1}$ (orange lines) as a function of electron temperature $T_e$. The solid and dot-dashed lines correspond to the numerical values and analytical forms (Eq.~\eqref{eq:num_conserv_mue} for $\mu_e$), respectively.}
%\label{fig:ce_and_mue}
%\end{figure}

\section{Discussion and outlook}\label{sec:discussion}

We have seen that the lower electron density along with the non-parabolicity distinguishes the electron and heat dynamics in ITO (and more generally, LEDD materials) from those in noble metals. In particular, we identified significant differences in the $e-ph$ interactions, a faster $e-e$ collision rate, a much stronger dependence of the TTM parameters on the electron temperature and a different overall heating and dynamics due to a lower electron heat capacity. The analytic expressions obtained for the TTM parameters allow an easy investigation of other LEDD materials.

We have also shown that the TTM matches remarkably well the dynamics of the effective electron temperature as well as of the phonon temperature. Nevertheless, the TTM has known limitations in noble metals; in particular, it assumes a-priori that the $e–e$ scattering is fast enough to establish a thermal distribution of electrons before significant energy is transferred to the phonons and cannot account for the accelerating rate of $e-ph$ collisions~\cite{non_eq_model_Lagendijk}. While this is a problematic assumption for noble metals, it was shown in~\cite{Wilson_Coh} that this assumption holds well in light metals such as Na, Cs, Rb, and K. In that respect, the condition of validity of the TTM in ITO does not strictly hold, yet, the faster $e-e$ collision rate makes it is closer to be satisfied in comparison to noble metals.

In order to improve upon the TTM, it is customary to add a dynamical equation for the total non-thermal energy (e.g.~\cite{non_eq_Fujimoto,non_eq_model_Carpene,nt_electrons,Dubi-Sivan,Dubi-Sivan-Faraday} for noble metals or~\cite{Boyd_Nat_Phot_2018} for ITO); this is usually done within the RTA, requiring a somewhat ambiguous choice of an energy-averaged decay coefficients of the non-thermal energy to the (thermal) electron and phonon subsystems. Whether such an improvement is necessary or not requires a rigorous consideration of the permittivity (or, e.g., reflectivity) dynamics; it might reveal differences between the thermal and non-thermal dynamics. This complicated task is left for a future paper. Nevertheless, even without such an analysis, we can already say that the decrease rate of the electron temperature  decay rate ($G_{e-ph}/C_e$) with the electron temperature implies that the internal thermalization process between these subsystems becomes slower with increasing illumination intensity. This explains the observation in~\cite{Sapienza_2022} of a slower thermalization of the reflectivity with increasing illumination intensities.

%Potentially, the most surprising consequence of the low density and low Fermi energy is that a suitable high intensity illumination can heat up the electrons even up to the Fermi temperature. This leads to an effective ``optical de-doping'' whereby a conducting LEDD material become a semiconductor exhibiting a negative chemical potential.

In this vein, the current work is a starting point for modelling the permittivity dynamics of the ITO and other LEDD materials, as well as the observed spectral broadening and fast switching, both for pulses which are a few 100's of femtoseconds long, as well as shorter ones. In this context, it would be of great interest to unravel the physical mechanism underlying the nonlinear optical response of ITO at increasingly high illumination intensities.

\acknowledgements I. W. Un and S. Sarkar contributed equally to this work. The authors were supported by Israel Science Foundation (ISF) grant (340/2020) and by Lower Saxony - Israel cooperation grant no. 76251-99-7/20 (ZN 3637). % Authors acknowledge E. Galiffi and R. Sapienza for valuable suggestions and discussions.

\appendix
\section{The $e-ph$ collision term and the $e-ph$ coupling coefficient}\label{app:e-ph-details}
\subsection{The phase-space argument for $e-ph$ scattering}\label{app:e-ph_phase_space}

In the main text, we claimed that the $e-ph$ collision term, the corresponding energy exchange rate and coupling coefficient can be dramatically overestimated in ITO if momentum conservation is not taken into consideration. However, since it is difficult to deduce how conservation of momentum is manifested in Eqs.~\eqref{eq:eph_polar_int}-\eqref{eq:invtau_ephn}, we provide below a detailed phase-space argument. 

\begin{figure}[h]
\centering
\includegraphics[width=1\textwidth]{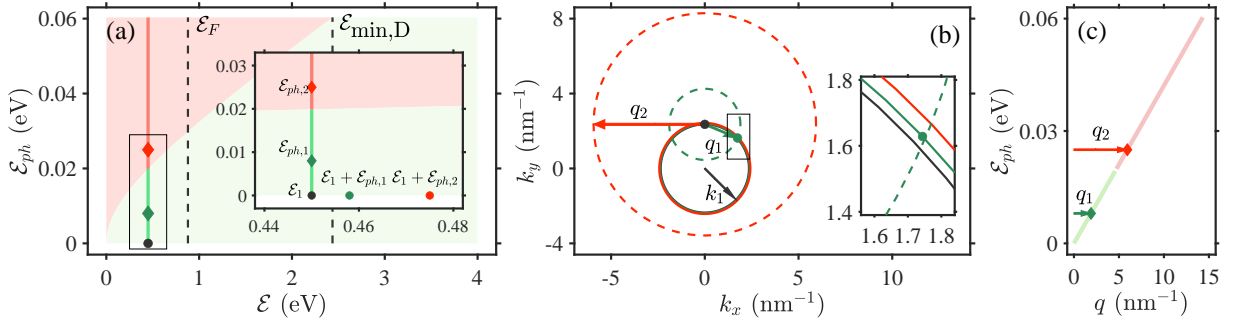}
\caption{(Color online) (a) Color plot showing the possibility of interaction between electrons with energy $\e$ and phonons with energy $\e_{ph}$: (green) interacting; (red) non-interacting. The inset shows that electrons with energy $\e_1$ can absorb a phonon with energy $\e_{ph,1}$ but cannot absorb a phonon with energy $\e_{ph,2}$. (b) The phase-space argument of the $e-ph$ interaction. The black, green and red circles respectively represent the states having energies $\e_1$, $\e_1 + \e_{ph,1}$ and $\e_1 + \e_{ph,2}$ on the $k_z = 0$ plane in the momentum space. The green (red) dashed circles represents the possible states that can be reached if an electron with energy $\e_1$ and momentum $(0,k_1,0)$ (black dot) absorbed a phonon with energy $\e_{ph,1}$ ($\e_{ph,2}$). The intersection between the green solid and dashed circles represents the final electron state of the scattering process. No intersection between the red solid and dashed circles indicates that no final state can be reached. (c) The linear energy-momentum relation of the acoustic phonon. The phonon states with energy $\e_{ph,1}$ ($\e_{ph,2}$) and momentum $q_1$ ($q_2$) are represented by the green (red) diamonds.}
\label{fig:dfdt_eph_mc}
\end{figure}

Let us consider an electron which initially has an energy $\e_1$ interacting with phonons having energy $\e_{ph,1}$ or $\e_{ph,2}$ as shown in green and red diamonds respectively in Figs.~\ref{fig:dfdt_eph_mc}(a) and (c). %In the momentum space, states having the same energy are represented by a sphere (a circle on the $k_z = 0$ plane) whose radius is equal to the magnitude of the momentum. 
Without loss of generality, we assume that the electron initially has momentum $(0,k_1 = \sqrt{2 m_e^\ast \e_1 (1 + C \e_1)}/\hbar,0)$ represented by the black dot in momentum space, see Fig.~\ref{fig:dfdt_eph_mc}(b). If the electron absorbs the phonon with energy $\e_{ph,1}$, %and momentum $q_1 = \e_{ph,1}/(\hbar v_[ph})$ 
the energy of the electron is changed to $\e_1 + \e_{ph,1}$ due to the conservation of energy. The possible final states can then be represented by a sphere (a circle on the $k_z = 0$ plane) whose radius is equal to the magnitude of the momentum $\sqrt{2 m_e^\ast (\e + \e_{ph,1})(1 + C(\e + \e_{ph,1}))}/\hbar$ represented by the green circle in Fig~\ref{fig:dfdt_eph_mc}(b). Meanwhile, the momentum of the electron is changed to $(0,k_1,0) + {\bf q}_1$ due to conservation of momentum, where ${\bf q}_1$ is the momentum of the absorbed phonon satisfying the linear energy-momentum dispersion $q_1 = \e_{ph}^1/(\hbar v_{ph})$ as shown in Fig~\ref{fig:dfdt_eph_mc}(c). The possible final states satisfying momentum conservation can then be represented by a sphere (a circle on the $k_z = 0$ plane) centered at $(0,k_1,0)$ and having a radius of $q_1$ (the green dashed circle in Fig~\ref{fig:dfdt_eph_mc}(b)) in momentum space.
%while the momentum of the electron is changed to $(0,k_1,0) + {\bf q}_1$ due to the conservation of momentum (represented by the red dashed circle in Fig~\ref{fig:dfdt_eph_mc}(b)).
Therefore, the true possible final electron states satisfying both energy and momentum conservation can then be identified by the intersection of these two spheres in the 3D momentum space (two circles on the $k_z = 0$ plane), as shown by the green dot in Fig.~\ref{fig:dfdt_eph_mc}(b).

%represents the  final electron states (the green dot in Fig.~\ref{fig:dfdt_eph_mc}(b)). 
Now, if the initial electron interacted with the phonon with energy $\e_{ph,2}$ and momentum ${\bf q}_2$, the final electron would have energy $\e_1 + \e_{ph,2}$ (represented by the red circle in Fig.~\ref{fig:dfdt_eph_mc}(b)) and momentum $(0,k_1,0) + {\bf q}_2$ (represented by the red dashed circle in Fig.~\ref{fig:dfdt_eph_mc}(b)). However, these two circles (two spheres in the 3D momentum space) do not intersect each other, indicating that no final state can be reached and thus the electrons with energy $\e_1$ do not interact with the phonons with energy $\e_{ph,2}$. 

Since the phonon energy is usually much smaller than the electron energy ($\e_D \ll \e_1$ in the above example), the initial and final energy ($\e_1$ and $\e_1 + \e_{ph,1}$, respectively) of the electron are very close to each other (that is why the black, green and red circles nearly overlap with each other in Fig.~\ref{fig:dfdt_eph_mc}(b)). However, the initial and final momentum of the electron can be very different and the magnitude of the momentum difference can range from 0 to $\sim 2 k_1$. Therefore, electrons with momentum $k_1$ (energy $\e_1$) interact only with phonons having momentum smaller than $\sim 2 k_1$ (energy smaller then $\sqrt{8m_e^\ast v_{ph}^2 \e_1(1 + C\e_1)}$, represented by the light green line in Fig.~\ref{fig:dfdt_eph_mc}(a) and (c))~\footnote{This explains why the polar integral in Eq.~\eqref{eq:eph_polar_int} is non-zero only when $q \lesssim 2 k$.}. This happens for electrons having momentum smaller than $q_D/2$ (or energy smaller than $\e_\text{min,D} = \dfrac{1}{2C}\left(\sqrt{1 + \dfrac{C\e_D^2}{2 m_e^\ast v_{ph}^2}} - 1\right)$), see Fig.~\ref{fig:dfdt_eph_mc}(a). %Conversely, this means that for electrons having momentum smaller than $q_D/2$ (energy smaller than $\e_\text{min,D} = \dfrac{1}{2C}\left(\sqrt{1 + \dfrac{C\e_D^2}{2 m_e^\ast v_{ph}^2}} - 1\right)$), the number of the phonon mode allowed for the electron-phonon interaction is restricted by the conservation of momentum, as shown in Fig.~\ref{fig:dfdt_eph_mc}(a). 
For ITO, $q_D/2 \approx 2 k_F $ is much larger than the Fermi momentum ($\e_\text{min,D} \approx 2.46$ eV is much larger than its Fermi energy), such that a substantial number of phonons are prohibited from interacting with electrons (see Fig.~\ref{fig:dfdt_eph_mc}(a)), resulting in a much smaller $e-ph$ coupling coefficient than that without accounting for momentum conservation in the electron-phonon collisions (see the comparison shown in Fig.~\ref{fig:Geph_ce_mu}(c)). %In contract, for noble metals, $\e_\text{min,D}$ ($\approx 0.46$ eV and $\approx 0.58$ eV for Au and Ag, respectively) is much smaller than their Fermi energy (5.53 eV and 5.49 eV for Au and Ag, respectively), so almost all electrons can interact with all phonons. 

\subsection{The function shape of the $e-ph$ collision term in the Boltzmann equation}\label{app:dfdt_eph_shape}
\begin{figure}[h]
\centering
\includegraphics[width=1\textwidth]{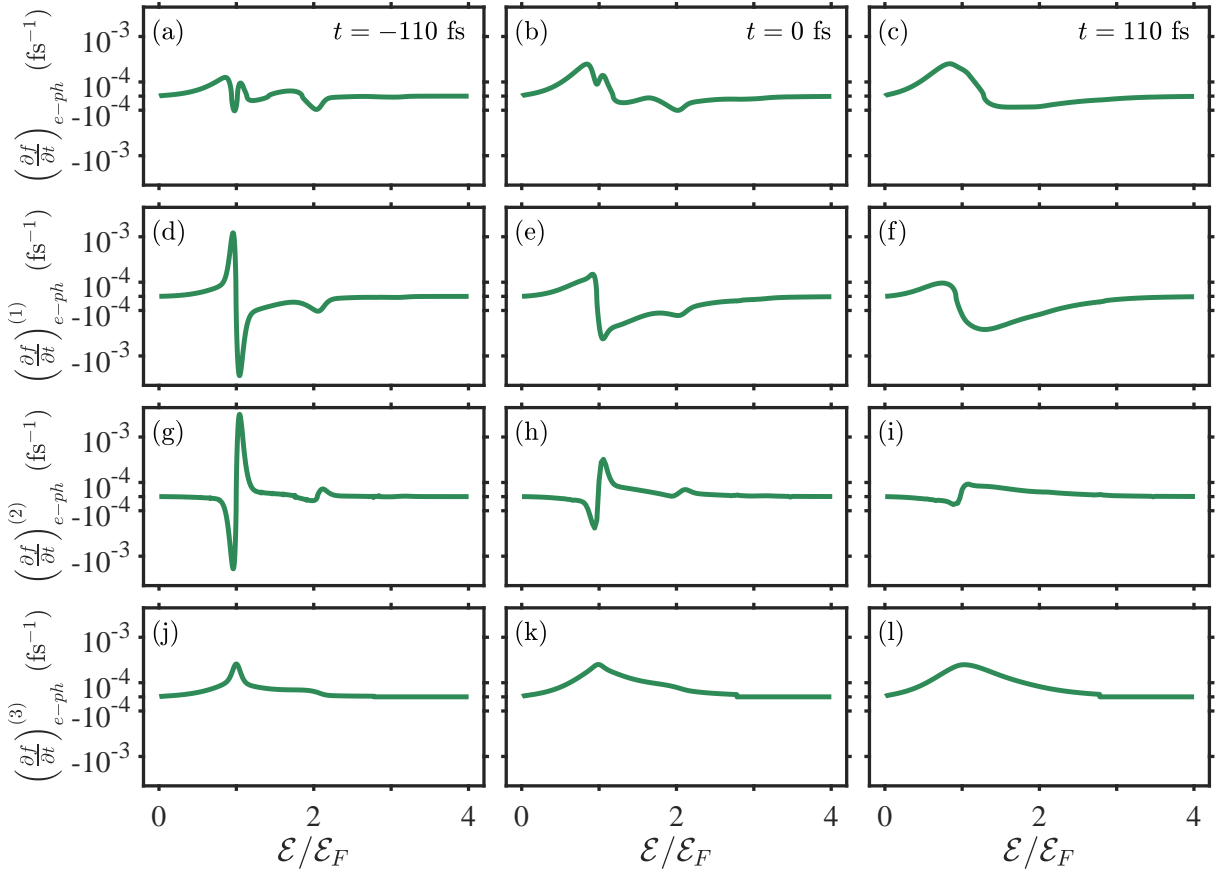}
\caption{(Color online) (a)-(c) the $e-ph$ collision term Eq.~\eqref{eq:dfEdt_eph} (the same as Fig.~\ref{fig:BE_terms}(c), (g) and (k)), (d)-(f) the first term, (g)-(i) the second term, and (j)-(l) the third term of Eq.~\eqref{eq:dfdt_eph_diff} following illumination of ITO by a short pulse. The left column shows the various terms at the FWHM \textit{before} the peak of the pulse ($t = -110$ fs), the middle column \textit{at the centre} of the pulse ($t =0$ fs), and the right column at the FWHM \textit{after} the peak of the pulse ($t = +110$ fs) as a function of $\mathcal{E}/\mathcal{E}_F$ and pulse intensity $I_\text{inc} = 2.5$ GW/cm$^2$. The $y$-axis is in symmetric log-scale.}
\label{fig:dfdt_eph}
\end{figure}

The conservation of momentum not only reduces the number of phonons available for $e-ph$ collisions, but also causes $\left(\dfrac{\partial f(\e)}{\partial t}\right)_{e-ph}$ to exhibit a very different shape from the shape characteristic of noble metals (see, e.g.,~\cite{vallee_nonequilib_2003,Stoll_review,Dubi-Sivan}), as shown in Fig.~\ref{fig:BE_terms}(c), (g), (k) and Fig.~\ref{fig:dfdt_eph}(a)-(c). To have a deeper understanding of this, we simplify Eq.~\eqref{eq:dfEdt_eph} by expanding its integrand in a power series in $\e_{ph}$. After some algebra, we find that $\left(\dfrac{\partial f(\e)}{\partial t}\right)_{e-ph}$ is dominated by three terms, namely,
\begin{multline}\label{eq:dfdt_eph_diff}
\left(\dfrac{\partial f}{\partial t}\right)_{e-ph} \approx \dfrac{D^2}{4\pi\rho(\hbar v_{ph})^4}\sqrt{\dfrac{m_e^\ast}{2\e(1+C\e)}}(1 + 2 C \e) \\
\Bigg[\dfrac{\left(\e_{ph}^\text{max}\right)^4}{4}\dfrac{\partial}{\partial \e} \left(f(\e)(1 - f(\e))\right) + \dfrac{\left(\e_{ph}^\text{max}\right)^5}{5} \dfrac{\partial^2}{\partial \e^2} f(\e) \\
+ 96 (m_e^\ast v_{ph}^2)^2 \e (1 + C\e) (1 + 2 C \e) f(\e)\left(1 - f(\e)\right) H(\e) H(\e_\text{min,D} - \e) \Bigg],
\end{multline}
where $\e_{ph}^\text{max}(\e) = \text{min}(\e_D,\sqrt{8 m_e^\ast v_{ph}^2\e(1 + C\e)})$ is the maximal energy of a phonon which can be absorbed/emitted by an electron with energy $\e$ (the boundary between the green and red regimes in Fig.~\ref{fig:dfdt_eph}(a)). The first term is proportional to the first derivative of $f(\e)(1 - f(\e))$ with respect to $\e$. The second term is proportional to the second derivative of $f(\e)$ with respect to $\e$. The third term is proportional to $f(\e)(1 - f(\e))$; it is non-zero only for $0<\e<\e_\text{min,D}$ (see the two Heaviside step functions) and it ensures electron number conservation in the $e-ph$ interaction, i.e., $\displaystyle\int\rho_e(\e)\left(\dfrac{\partial f}{\partial t}\right)_{e-ph}d\e = 0$. For noble metals, $\e_\text{min,D}$ is much smaller than Fermi energy such that $\e_{ph}^\text{max}(\e) = \e_D$ for most of the electrons and the contribution from the third term becomes negligible. In this case, Eq.~\eqref{eq:dfdt_eph_diff} reduces to the usual differential form of the $e-ph$ collision~\cite{vallee_nonequilib_2003}.

Fig.~\ref{fig:dfdt_eph}(d)-(l) show the three terms in Eq.~\eqref{eq:dfdt_eph_diff} as a function of electron energy at three different time, before ($t=-110$ fs), at the centre of ($t=0$ fs) and after ($t=+110$ fs) the peak of the pulse. The shape of these three terms can be explained by the (smeared) multi-stair-step structure of $f(\e)$ (see Fig.~\ref{fig:BE_terms}(a)). Since the third term in Eq.~\eqref{eq:dfdt_eph_diff} has the simplest form (proportional to $\sim f(\e)(1-f(\e))$) and the first term is proportional to the first derivative of $f(\e)(1-f(\e))$, we start with explaining the function shape of the third term; then the first term; and lastly the second term.

To explain the function shape of the third term, we first look at the multi-stair-step shape of $f(\e)$ at $t = -110$ fs as shown in Fig.~\ref{fig:BE_terms}(a). The step of $f(\e)$ near $\e_F$ is due to the Fermi–Dirac nature. Around this step, $f(\e)$ changes from $\sim 1$ to $\sim 0$ while $1-f(\e)$ changes from $\sim 0$ to $\sim 1$. This leads to a peak in the third term of Eq.~\eqref{eq:dfdt_eph_diff} ($\sim f(\e)(1 - f(\e))$) near the Fermi energy, see Fig.~\ref{fig:dfdt_eph}(j). The small step of $f(\e)$ near $\e_F+\hbar\omega$ is created by the photon absorption (the non-thermal shoulder). Around this step, $f(\e)$ changes from $\sim 10^{-2}$ to $\sim 10^{-4}$, while $1-f(\e)$ is nearly equal to 1. This causes the third term of Eq.~\eqref{eq:dfdt_eph_diff} ($\sim f(\e)(1 - f(\e))$) to have a step-like shape near $\e_F + \hbar\omega$ (instead of a peak), see Fig.~\ref{fig:dfdt_eph}(j). For $t = 0$ and $t = 110$ fs, due to the $e-e$ collision, the electron distribution is smeared out (see Fig.~\ref{fig:BE_terms}(e) and~(i)). This also smears out the peak and the step of the third term in Eq.~\eqref{eq:dfdt_eph_diff}, see Fig.~\ref{fig:dfdt_eph}(k) and (l). The peak and the step of $f(\e)(1-f(\e))$ then respectively lead to a Lorentzian dispersion shape near the Fermi-energy and a dip near $\e_F+\hbar\omega$ in the first term of Eq.~\eqref{eq:dfdt_eph_diff} since it is proportional to the first derivative of $f(\e)(1-f(\e))$, see Fig.~\ref{fig:dfdt_eph}(d)-(f). Finally, since the second term of Eq.~\eqref{eq:dfdt_eph_diff} is proportional to the second derivative of $f(\e)$ with respect to $\e$, it has a Lorentzian dispersion shape near $\e_F$ and $\e_F+\hbar\omega$, see Fig.~\ref{fig:dfdt_eph}(g)-(i). 

Both the first and the second terms have a Lorentzian dispersion shape near $\e_F$ but they are in opposite sign, the combination of these two terms thus also has a Lorentzian dispersion shape. Combining this with the peak from the third term results in the complicated shape of $\left(\dfrac{\partial f}{\partial t}\right)_{e-ph}$ near the Fermi energy. Finally, the dip of $\left(\dfrac{\partial f}{\partial t}\right)_{e-ph}$ near $\e_F+\hbar\omega$ is mainly contributed from the first term in Eq.~\eqref{eq:dfdt_eph_diff}, see Fig.~\ref{fig:dfdt_eph}(a)-(c). 

\subsection{The analytical expression of the $e-ph$ coupling coefficient $G_{e-ph}$}\label{app:G_eph_analy_expres}
In this Appendix, we provide the analytical expression of the $T_e$-dependent $e-ph$ coupling coefficient $G_{e-ph}$ (Eq.~\eqref{eq:G_eph_int}). We follow the procedure mentioned in the main text, exchange the integral order, and separate the right-hand side of Eq.~\eqref{eq:G_eph_int} into two terms, $G_{e-ph}(T_e) = G^{(0)}_{e-ph}(T_e) + G^{(1)}_{e-ph}(T_e)$,
\begin{multline}\label{eq:Gephn0_NPC}
G^{(0)}_{e-ph}(T_e) \approx \dfrac{D^2}{4\pi^3\rho(\hbar v_{ph})^4}\dfrac{m_e^{\ast2}k_B}{\hbar^3}
\int_0^{\e_D} d\e_{ph}(\e_{ph})^3\int_0^{\infty}d\e \dfrac{\e(1 + 2 C \e)}{(2 k_B T_e)^2}\\ \Big[(1 + 2 C \e)\sech^2\left(\dfrac{\e - \mu}{2 k_B T_e}\right)\tanh\left(\dfrac{\e - \mu}{2k_B T_e}\right) - 4 C k_B T_e\sech^2\left(\dfrac{\e - \mu}{2k_B T_e}\right)\Big]
\end{multline}
and
\begin{multline}\label{eq:Gephn1_NPC}
G^{(1)}_{e-ph}(T_e) \approx - \dfrac{D^2}{4\pi^3\rho(\hbar v_{ph})^4}\dfrac{m_e^{\ast2} k_B}{\hbar^3} \Bigg\{\int_0^{\e_D} d\e_{ph}(\e_{ph})^3 \int_0^{\e^\text{min}} \dfrac{\e(1 + 2 C \e)}{(2 k_B T_e)^2} d\e
\\
\Big[(1 + 2 C \e)\sech^2 \left(\dfrac{\e-\mu}{2 k_B T_e}\right) \tanh\left(\dfrac{\e - \mu}{2 k_B T_e}\right) - 4 C k_B T_e \sech^2\left(\dfrac{\e - \mu}{2 k_B T_e}\right)\Big] \\
+ \int_0^{\e_D} d\e_{ph}(\e_{ph})^3 \dfrac{\e^\text{min}}{4 k_B T_e} (1 + 2 C \e^\text{min})^2 \sech^2\left(\dfrac{\e^\text{min} - \mu}{2 k_B T_e}\right) \Bigg\},
\end{multline}
where $\e^\text{min}(\e_{ph}) = \dfrac{1}{2 C}\left(\sqrt{1 + \dfrac{C(\e_{ph})^2}{2 m_e^\ast v_{ph}^2}} - 1\right)$ is the minimum energy of an electron which can absorb/emit a phonon with energy $\e_{ph}$. Next, we change the variables from $\e$ to $x = \dfrac{\e - \mu}{2 k_B T_e}$ and from $\e_{ph}$ to $x^\text{min} = \dfrac{\e^\text{min} - \mu}{2 k_B T_e}$ so that $\displaystyle\int_0^\infty d\e = \displaystyle \int_{x_0}^\infty(2 k_B T_e)dx$, $\displaystyle\int_0^{\e^\text{min}}d\e = \displaystyle\int_{x_0}^{x^\text{min}}(2 k_B T_e) dx$ and 
$\displaystyle\int_0^{\e_D} d\e_{ph} \e_{ph} = \dfrac{1}{2}(8 m_e^\ast v_{ph}^2)(2k_BT_e)\displaystyle\int_{x_0}^{x_D}dx^\text{min}(1 + 4Ck_BT_e(x^\text{min} - x_0))$, where $x_0 = -\dfrac{\mu}{k_B T_e}$, $x_D = \dfrac{\e_\text{min,D} - \mu}{k_B T_e}$ and $\e_\text{min,D} = \dfrac{1}{2C}\left(\sqrt{1 + \dfrac{C\e_D^2}{2 m_e^\ast v_{ph}^2}} - 1\right)$. Then, Eqs.~\eqref{eq:Gephn0_NPC} and~\eqref{eq:Gephn1_NPC} becomes 
\begin{multline}\label{eq:Gephn0_int}
G^{(0)}_\text{e-ph} \approx \dfrac{D^2}{4\pi^3\rho(\hbar v_{ph})^4}\dfrac{m_e^{\ast 2}k_B}{\hbar^3} \dfrac{\e_D^4}{4} \int_{x_0}^\infty dx \Big[ (x - x_0)(1 + 4 C k_B T_e(x - x_0))^2\sech^2(x)\tanh{x} \\
- 4 C k_B T_e(x - x_0)(1 + 4 C k_B T_e(x - x_0)) \sech^2(x) \Big],
\end{multline}
and
\begin{multline}\label{eq:Gephn1_int}
G^{(1)}_\text{e-ph} \approx -\dfrac{D^2}{4\pi^3\rho(\hbar v_{ph})^4}\dfrac{m_e^{\ast 2} k_B}{\hbar^3}(8 m_e^\ast v_{ph}^2)^2(2 k_B T_e)^2  \\
\Bigg\{\dfrac{1}{2}\int_{x_0}^{x_D} dx^\text{min}(x^\text{min} - x_0)(1 + 2 C k_B T_e(x^\text{min} - x_0))(1 + 4 C k_B T_e(x^\text{min} - x_0))
\\
\int_{x_0}^{x^\text{min}} dx \Big[ (x - x_0)(1 + 4 C k_B T_e(x - x_0))^2 \sech^2(x)\tanh{x} \\
- 4 C k_B T_e(x - x_0)(1 + 4 C k_B T_e(x - x_0))\sech^2(x) \Big] \\
+ \dfrac{1}{4}\int_{x_0}^{x_D} dx^\text{min}
(x^\text{min} - x_0)^2(1 + 2 C k_B T_e(x^\text{min} - x_0)) \\
(1 + 4 C k_B T_e(x^\text{min} - x_0))^3 \sech^2(x^\text{min})\Bigg\}.
\end{multline}
To evaluate Eqs.~\eqref{eq:Gephn0_int} and~\eqref{eq:Gephn1_int} analytically, we define
\begin{align}
g_n(x) \equiv \dfrac{x^n}{n!} - \ln2\dfrac{x^{n-1}}{(n-1)!} - \left(-\dfrac{1}{2}\right)^{n-1} \text{Li}_n(-e^{-2x}),\ n \in \mathbb{Z}
\end{align}
One can verify that $g_n(x) = \dfrac{dg_{n+1}(x)}{dx}$, $g_{-2}(x) = - 2 \sech^2(x)\tanh(x)$ and $g_{-1}(x) = \sech^2(x)$. The integral in Eqs.~\eqref{eq:Gephn0_int} and~\eqref{eq:Gephn1_int} can thus be expressed using 
\begin{multline}\label{eq:intxmgn}
\int_{x_0}^{x^\text{min}}dx(x - x_0)^m g_n(x) = \\ \sum_{r=0}^m \left[\dfrac{(-1)^r m!}{(m-r)!}(x^\text{min} - x_0)^{m-r} g_{n+r+1}(x^\text{min})\right] - (-1)^m g_{n+m+1}(x_0)
\end{multline}
and
\begin{multline}\label{eq:intxkxmgn}
\int_{x_0}^{x_D}dx^\text{min}(x^\text{min}-x_0)^k \int_{x_0}^{x^\text{min}}dx(x-x_0)^m g_n(x)  = \\
\sum_{r=0}^{m}\sum_{s=0}^{m+k-r}\left[\dfrac{(-1)^{r+s}m!}{(m-r)!}\dfrac{(m + k - r)!}{(m + k - r - s)!}(x_D - x_0)^{m + k - r - s} g_{n + r + s + 2}(x_D)\right]\\
- \sum_{r=0}^{m}\left[\dfrac{(-1)^{m + k}m!(m + k - r)!}{(m - r)!}g_{n + m + k + 2}(x_0)\right] - (-1)^m\dfrac{(x_D - x_0)^{k+1}}{(k + 1)!}g_{n + m + 1}(x_0),
\end{multline}
where $m$ and $k$ are positive integers. 

One can then obtain the full analytical expression of $G_{e-ph}(T_e)$ by substituting Eqs.~\eqref{eq:intxmgn}-\eqref{eq:intxkxmgn} into Eq.~\eqref{eq:Gephn0_int} and~\eqref{eq:Gephn1_int}. Since the full expression is too long, we shall not present it here. At low electron temperatures, i.e., $k_B T_e \ll \mu$, we have $e^{-\frac{\mu}{k_B T_e}} \rightarrow0$, $\ln\left(1+e^{\frac{\mu}{k_B T_e}}\right)\rightarrow\dfrac{\mu}{k_B T_e}$ and $\text{Li}_2\left(-e^{-\frac{\mu}{k_BT_e}}\right)\rightarrow-2\left(\dfrac{\mu}{2k_BT_e}\right)^2$, then the sum of Eq.~\eqref{eq:Gephn0_int} and Eq.~\eqref{eq:Gephn1_int} reproduces  Eq.~\eqref{eq:G_eph_int_0K}.

%The advantage of this expression over the seeming simpler numerical evaluation is that it is suitable for any intrinsic parameters of ITO (or other TCOs) and that it can be directly used for TTM calculations. 

\bibliography{my_bib}
\end{document}